\RequirePackage{fix-cm}
\documentclass[twocolumn]{svjour3}          
\smartqed  

\usepackage{amsmath}
\usepackage{graphicx}
\usepackage[T1]{fontenc}
\usepackage[utf8,latin1]{inputenc}
\usepackage{epsfig}
\usepackage{txfonts}
\usepackage{amssymb}
\usepackage{siunitx}
\usepackage{natbib}
\usepackage{hyperref}

\usepackage{url}
\usepackage{color}
\usepackage{bm}
\usepackage{sgl-commands}
\journalname{}

\DeclareMathOperator{\sgn}{sgn}
\newcommand\xilen{\boldsymbol{\xi}}
\newcommand\etalen{\boldsymbol{\eta}}
\newcommand\bx{\boldsymbol{x}}
\newcommand\mchi{\raise2pt\hbox{$\chi$}}
\newcommand{\Tds}{T_\mathrm{ds}}
\newcommand{\rhocr}{\rho_{\rm cr}}

\definecolor{dgreen}{rgb}{0,0.5,0}

\begin{document}

\title{Essentials of strong gravitational lensing}
\subtitle{}
\date{}

\titlerunning{}

\author{Prasenjit Saha \and Dominique Sluse \and Jenny Wagner \and
  Liliya L.~R.~Williams}

\authorrunning{P. Saha, D. Sluse, J. Wagner, and L.L.R. Williams}

\institute{P.~Saha \at
  Physik-Institut, University of Zurich,
  Winterthurerstrasse 190, 8\ 057 Zurich, Switzerland \\
  ORCID: https://orcid.org/0000-0003-0136-2153 \\
  \email{psaha@physik.uzh.ch} \\
  Corresponding author
  \and
  D.~Sluse \at
  STAR Institute, Quartier Agora All\'{e}e du six Ao\^{u}t,
  19c B-4000 Li\`{e}ge, Belgium \\
  ORCID: https://orcid.org/0000-0001-6116-2095 \\
  \email{dsluse@uliege.be}
  \and
  J.~Wagner \at
  Bahamas Advanced Study Institute and Conferences, 
  4A Ocean Heights, Hill View Circle, Stella Maris, Long Island, The Bahamas
  ORCID: https://orcid.org/0000-0002-4999-3838 \\
  \email{thegravitygrinch@gmail.com}
  \and
  L.~L.~R.~Williams \at
  School of Physics and Astronomy, University of Minnesota,
  116 Church Street, Minneapolis, MN 55455, USA \\
  ORCID: https://orcid.org/0000-0002-6039-8706 \\
  \email{llrw@umn.edu}
 }

\maketitle

\begin{abstract}
Of order one in $10^3$ quasars and high-redshift galaxies appears in
the sky as multiple images as a result of gravitational lensing by
unrelated galaxies and clusters that happen to be in the foreground.
While the basic phenomenon is a straightforward consequence of general
relativity, there are many non-obvious consequences that make
multiple-image lensing systems (aka strong gravitational lenses)
remarkable astrophysical probes in several different ways.  This
article is an introduction to the essential concepts and terminology
in this area, emphasizing physical insight.  The key construct is the
Fermat potential or arrival-time surface: from it the standard lens
equation, and the notions of image parities, magnification, critical
curves, caustics, and degeneracies all follow.  The advantages and
limitations of the usual simplifying assumptions (geometrical optics,
small angles, weak fields, thin lenses) are noted, and to the extent
possible briefly, it is explained how to go beyond these. Some less
well-known ideas are discussed at length: arguments using wavefronts
show that much of the theory carries over unchanged to the regime of
strong gravitational fields; saddle-point contours explain how even
the most complicated image configurations are made up of just two
ingredients.  Orders of magnitude, and the question of why
strong lensing is most common for objects at cosmological distance, are also
discussed.  The challenges of lens modeling, and diverse strategies
developed to overcome them, are discussed in general terms, without many technical details.
\end{abstract}

\newpage
\tableofcontents


\section{The general picture}

Some time in centuries past, pieces of glass in shapes resembling
lentil seeds ({\em Lens culinaris}) came to be known as lenses.  The
name remains with us: the optical elements developed for lighthouses
in the 19th century are still called lenses, even though they look
more like giant pineapples.  A lens, then, is anything that transmits
light with some non-trivial optical path or light travel time, causing
light deflection. If an interesting optical path is created by a
gravitational field, meaning the spacetime metric, we have a
gravitational lens.

\begin{figure}
\hbox to \hsize{\hss
\includegraphics[width=\hsize]{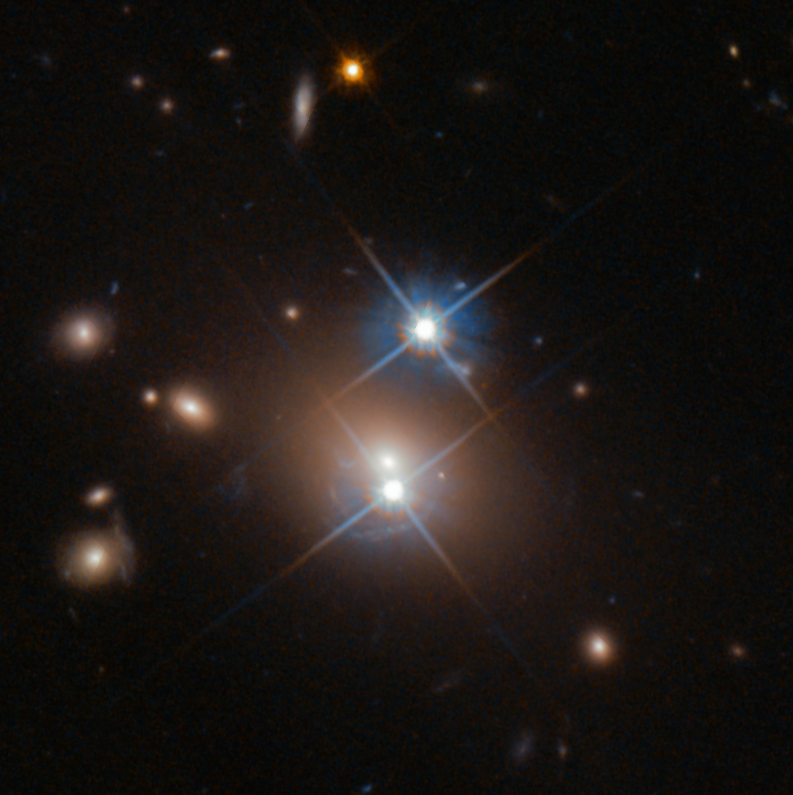}\hss}
\caption{The archetype of strong gravitational lensing: Q0957+561.
  The two bright bluish objects with diffraction spikes are lensed
  images of a quasar at redshift $z=1.41$.  The elliptical galaxy near
  the lower quasar image, and the other galaxies in the field are part
  of a cluster at $z=0.36$, which together form the gravitational
  lens.  The field shown is $30''\times30''$ with North up and East to the left.
\label{fig:archetype}}
\end{figure}

Light deflection in accordance with Einstein gravity was measured in
1919, and theorists have written about gravitational lensing for even
longer\footnote{For the fascinating scientific history, often at odds
with textbook narratives, see \cite{bib1:Valls2006} and
\cite{Kennefick2012}.}, but it was in the Einstein centennial year of
1979, with the discovery of the ``double quasar'' Q0957+561 by
\cite{bib1:Walsh1979}, that gravitational lensing became part of
astronomy.  A beautiful image of this system is shown in
Fig.~\ref{fig:archetype}.

In the decades since 1979 more than a thousand multiply-imaging
gravitational-lens systems comparable to Q0957+561 have been
discovered, and the discovery rate is accelerating.  This article will
introduce the essential theory needed to study such systems, which are
also called ``strong-lensing'' systems.  The latter term contrasts the
phenomenon of multiple image with weak-lensing, where there is light
deflection and image distortion but not multiple images --- such as in
\cite{2021A&A...645A.105G} on weak lensing by large-scale structure,
or \cite{2017PhRvD..96j4030C} on weak lensing in the solar system.
Note, however, that strong lensing still occurs in weak gravitational
fields --- we will not consider light propagation near event horizons
\citep[as in the well-known images from][]{2022ApJ...930L..17E}.  Some
of the theory is also applicable to gravitational microlensing by
stars in or near the Milky way \citep[e.g.,][]{2020ApJS..249...16M},
but mainly we will be concerned with lensing by galaxies and clusters
at cosmological distances.

\subsection{Light deflection}

The practice of strong gravitational lensing involves some simplifying
assumptions.

First, geometrical optics is assumed for light sources and the
gravitational-lensing processes.  Wave optics is essential at the
telescope, but we do not consider interference between multiple
images.  Appendix~\ref{sec:app-waves} briefly discusses how the
standard formalism can be extended to wave optics, but for the rest of
the present article, light means rays.

Second, gravitational fields are assumed to be weak.  As a
consequence, light rays are deflected from straight lines only by very
small angles.  Furthermore, the contributions of many masses can be
simply added (or integrated over) without the need to solve Einstein's
field equations.  Thus, lensing is treated as a first-order
perturbation from light travelling along straight lines.

A third assumption is that light deflection occurs over regions much
smaller than the distances between light source, lensing mass, and
observer.  This assumption is expressed in the notions of a
\textit{lens plane} and a \textit{source plane}.  It is not meant that
sources and lenses are flat, but rather that the gradual deflection of
light through a gravitational field is integrated along the line of
sight through the comparatively small active region, and approximated
as a discrete deflection.  Distances transverse to the line of sight,
however, are treated with respect, and not unceremoniously integrated
out.

The well-known expression
\begin{equation}
  \frac{4GM_\odot}{c^2R_\odot} = 1.75'' \simeq \SI{1e-5}{\radian}
  \label{eq:solaralpha}
\end{equation}
for gravitational light deflection at the rim of the Sun exemplifies
all three assumptions: (i)~geometrical optics is understood, (ii)~the
metric perturbations are of the same order as the deflection in
radians, and thus small, and (iii)~the Sun is approximated as a
flat deflector in the sky.  Light deflections by galaxies are
coincidentally of the same order as that by the Sun, while clusters of
galaxies produce deflections an order of magnitude larger.  Thus, the
light deflections that will concern us in practice are
$\lesssim\SI{1e-4}{\radian}$ and entirely in the regime of weak
gravitational fields.

Introductions to gravitational lensing
\citep[e.g.,][]{2021LNP...956.....M} conventionally start by
specializing to the regime of geometrical optics, weak fields, and a
planar lens.  Some of the essential concepts, however, are also valid
for strong fields.  So let us start a little outside the usual comfort
zone, with an example of lensing by a strong gravitational field.

\begin{figure}
\hbox to \hsize{\hss
\includegraphics[width=\hsize]{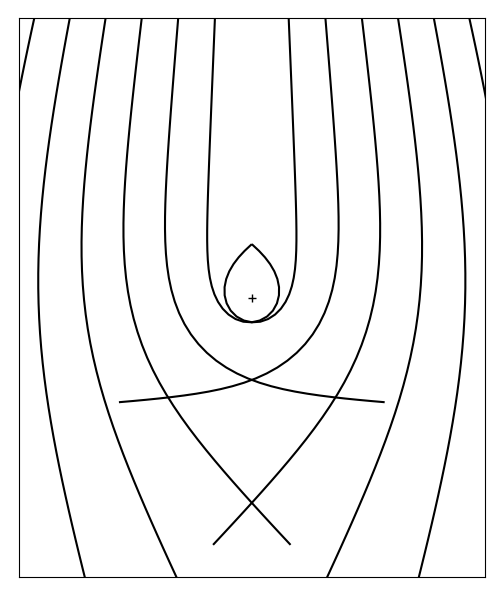}\hss}
\caption{A representation of the deflection of light rays.  The source
  is far above, outside the figure.  There is a strong gravitational
  field near the center of the figure (marked with a $+$), which
  produces large deflections (even $>180^\circ$).  Farther from the
  center, the field is weaker and the deflections smaller, though still
  much larger than the real systems we are considering.
\label{fig:rays}}
\end{figure}

\figref{fig:rays} shows light deflection through one of the simplest curved
spacetimes.  The metric (see Appendix~\ref{sec:app-exmods} for details)
is similar to a Schwarzschild metric, but has no event horizon
(which, though fascinating, would take too long to discuss here).  
Light rays
originate from a point source high above the figure, and are gradually
diverging as they approach the strong field in the middle of the figure,
and then the lens pulls the rays together.  There is no focal point,
though there is a sort of focal line along which rays converge
symmetrically from paths on opposite sides of the lens.
Two such ray intersections on the focal line are shown in the figure.
Observers located there will observe a ring of light, provided the
lens is circularly symmetric.  There are also asymmetric ray
intersections (two of which are indicated) in the lower part of the
figure, not on the focal line.  Observers located there will see the
source light coming from two directions in their sky.

The rays near the middle of \figref{fig:rays} undergo very large
deflections, even more than $180^\circ$.  Large deflections occur in
strong gravitational fields near neutron stars and black holes,
leading to the exotic multiple imaging now observed near the M87 black
hole \citep{2022ApJ...935...61B}.  It does not, however, occur in the
systems we will consider.

Lensing theory conventionally uses observer-centric coordinates.  To a
given observer, a source appears to be located along a direction (say
$\tang$), whereas its true location is along some different direction
(say $\bang$).  The directions $\tang$ and $\bang$ are conventionally
expressed as two-dimensional angles on the sky, measured in radians or
arc-seconds.  The mapping relating $\tang$ and $\bang$ is called the
lens equation, and is conventionally expressed as
\begin{equation}
  \bang = \tang - \aang(\tang)
\label{eq:lenseq}
\end{equation}
with $\aang(\tang)$ having the interpretation of an apparent
deflection angle.
Adding angles in this way, one is tacitly assuming that the sky has
been mapped to a plane.  For large deflections we would need to
specify the sky projection, but if all angles are small, at any time,
only a very small patch of sky is relevant, and angles can be simply
added.

The convention of using observer-centric coordinates seems peculiar to
gravitational lensing.  In optics, it is usual to put the origin at
the lens.  Even within astronomy, the study of planetary atmospheres
as lenses for occulted stars \citep[see e.g.,][]{1996AREPS..24...89E}
uses lens-centric coordinates.  Observer-centric coordinates are,
however, a good idea, because some useful facts get built into the
notation.  As we can see from the form of the lens equation
(\ref{eq:lenseq}), the mapping from image plane $\tang$ to source
plane $\bang$ is single valued, but the inverse can be multiple
valued.  Thus each image corresponds to a unique source, whereas each
source can have multiple images.  Ray-tracing using the lens equation
(\ref{eq:lenseq}) using your favorite deflection field $\aang(\tang)$
is easy and can be entertaining.\footnote{Try visiting
$$ \hbox{\url{https://phdenzel.github.io/streaming-lens/}} $$ on a
  device with a camera.}  Each pixel in the image plane is given
the brightness of the corresponding pixel in the source plane.  (Since
$\bang$ is a single-valued function of $\tang$, but not the other way
round, ray tracing is usually done backwards, from the observer's sky
to the source.)
Note that ray tracing in this way automatically preserves surface
brightness (photons per steradian).  A patch of uniform brightness on
the source stays the same uniform brightness when lensed, but it can
change in size and shape.

The conventional minus sign for the deflection in
Eq.~(\ref{eq:lenseq}) means that $\aang(\tang)$ points away from the
deflector rather towards it.  This is an example of a well-known
principle in astrophysics,\footnote{Attributed to J.~Binney.}
which is that you first think of how a
rational person would do it, and then you \textit{change the
  sign}.

\begin{figure}
\hbox to \hsize{\hss
\includegraphics[width=\hsize]{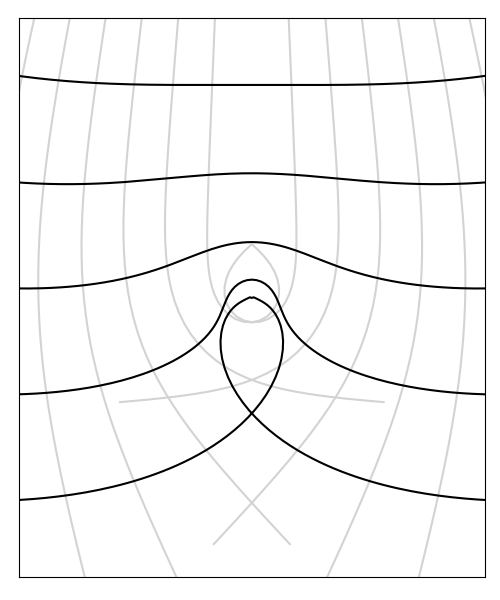}\hss}
\caption{Wavefronts corresponding to the light rays in
  \figref{fig:rays}.  (The source is above outside the figure.) Light gray curves are taken over from that
  figure.  Each of the black curves is the endpoint of all the rays at
  a particular time.  The small loop is the result of very large
  deflections and would only occur in a strong gravitational field.
\label{fig:wavefront}}
\end{figure}

\subsection{Wavefronts and Fermat's principle}

The lens equation (\ref{eq:lenseq}) would actually be too general for
lenses, if we let $\aang(\tang)$ be arbitrary.  This is because you
cannot send light any way you like using just lenses --- for that you
would need optical fibers.  Imagine a cable consisting of a bundle of
twisting optical fibers, and a signal input simultaneously into all
the fibers.  The overall direction of travel of the light signal along
the cable will be different from the direction of travel within
individual fibers.  Lenses cannot make light travel like this. In
lensing, the surface or front described by a bundle of light rays as
light travels is always normal to the rays themselves.  This front is
known as the wavefront.  (Despite the name, the wavefront is
well-defined in geometrical optics.  If the source is small but not
effectively a point, the wavefront will be wiggly at the wavelength
scale, and will not produce interference.)

\figref{fig:wavefront} is a representation of the wavefront
corresponding to the rays in \figref{fig:rays}.  Here, we imagine the
source emitting an omni-directional light flash, and the wavefront
shows how far that light flash has travelled in a given time in
different directions.  Light rays are simply the local normals to the
wavefront.  Notice that the lens delays the wavefront --- drastically
in the strong-field region, less in other places.  Most interesting
for strong gravitational lensing, some observers will be crossed by
the wavefront more than once, at different times and from different
directions.

\begin{figure}
\hbox to \hsize{\hss
\includegraphics[width=\hsize]{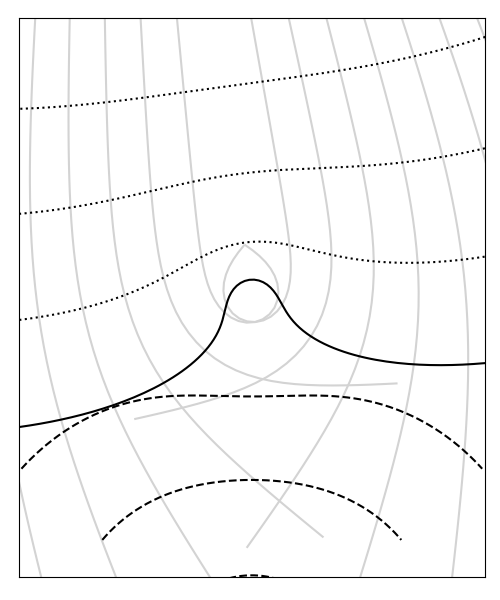}\hss}
\caption{Modified from \figref{fig:wavefront}.  The source (outside the figure) has been
  moved slightly to the left.  The wavefront proceeds as the dotted
  curves and is then frozen at the solid black curve.  Then a second
  backward wavefront (dashed) emerges from the observer on the ground.
  Each of the dashed curves is the endpoint at a particular time of
  rays emanating from the observer, but those rays are not shown.
\label{fig:twowf}}
\end{figure}

Let us now modify the preceding to the scenario shown in
\figref{fig:twowf}.  Let us pivot the source (located above and
outside the figure) by a small angle $\bang$ about the lens.  The
incoming wavefront is then inclined by $\bang$.  After the wavefront
has passed the lens and been delayed by it, we freeze the wavefront.
Then we imagine a second backwards wavefront, which emanates from an
observer at the bottom of the figure.  Any part of this second
wavefront can be labelled by the angle $\tang$ on the observer's sky.
It is like the observer's sky rising up to meet the frozen wavefront.
Let $t(\tang)$ be the time at which a point $\tang$ on the rising-sky
wavefront meets the original frozen wavefront.  Let us now concentrate
on any places where the two wavefronts meet tangentially.  These
correspond to $\nabla t(\tang)=0$.  At these places the normals to the
two wavefronts are the same.  Since normals to the wavefront are light
rays, $\nabla t(\tang)=0$ corresponds to light rays travelling out
from the source with a wavefront and then down the sky to the
observer.  Moreover, $t(\tang)$ at these sky locations is the light
travel time, apart from an additive constant.
This is Fermat's principle in gravitational lensing.

The two-wavefront picture of lensing \citep[due
  to][]{bib1:Nityananda1990} is remarkable in that, from the basic
postulate that light rays are normal to a wavefront, it leads
inexorably to Fermat's principle and the notion of an abstract surface
$t(\tang)$, even with large deflections and strong fields.  We assumed
a stationary lens, but in fact Fermat's principle in terms of an
arrival time can be extended to arbitrary time-varying gravitational
fields in general relativity \citep{1992PhRvD..45.3862N}.  In the
following we will specialize to small deflections, but it is worth
emphasizing that the notion of the arrival-time surface $t(\tang)$ and
its consequences have much wider applicability.

\subsection{Geometrical and gravitational time delays}\label{subsec:geomgrav}

Let us now turn \figref{fig:twowf} into equations, under the following
simplifying assumptions.
\begin{itemize}
\item All angles are taken to be small, in the sense of
  $\sin\theta\approx\theta$ and $\cos\theta\approx1-\frac12\theta^2$.
\item \figref{fig:twowf} has the source far away from the lens.  We
  now assume that the source is
  at infinity.  (We will relax this assumption later.)
  As before, on the observer's sky the source is located
  at location $\bang$.
\item The gravitational field is assumed localized to the neighborhood
  of the lens.  The backward wavefront is therefore spherical.
\end{itemize}
Under these assumptions, the frozen wavefront just after passing the
lens is a slightly warped plane inclined by $\bang$, while the dashed
backward wavefront is $\propto\theta^2$.

Without any lensing mass, the frozen wavefront is simply a plane, and
the time taken for backward wavefront to reach it follows from
geometry as
\begin{equation}
  t_{\rm geom}(\tang) = \frac Dc \, \left(
  {\textstyle\frac12} |\tang\,|^2
  - \tang\cdot\bang \,\right) + \hbox{const} \;,
\end{equation}
where $D$ is the shortest distance from the observer to the frozen
wavefront.  Since $\bang$ is constant, we can equivalently write
\begin{equation}
  t_{\rm geom}(\tang) = \frac D{2c} |\tang-\bang \,|^2 \;.
  + \hbox{const}
\end{equation}

Adding a lensing mass, the frozen wavefront gets warped, and the time taken by the backward wavefront to reach it
changes by say $t_{\rm grav}(\tang)$.  The light travel time, up to a
constant, would then be
\begin{equation}
t(\tang) = t_{\rm grav}(\tang) + t_{\rm geom}(\tang) \;.
\end{equation}
The form of $t_{\rm grav}(\tang)$ is not obvious, but let us put in the
ansatz
\begin{equation}
  t_{\rm grav}(\tang) = - \frac G{c^3}
  \sum_i M_i \ln|\tang-\tang_i|
\label{eq:refsdal-shapiro}
\end{equation}
pending verification that it leads to the correct deflection angle.
Note here, that, since the angles are all small, the logarithm will
always be negative, and because of the minus sign the contribution of
the mass term will always be positive.  Light rays will reach the
observer and images will appear where the gradient of $\nabla
t(\tang)=0$.  Working out the gradient we have
\begin{equation}
  \bang = \tang - \frac{4G}{c^2D}
  \sum_i M_i \, \frac{\tang-\tang_i}{|\tang-\tang_i|\rlap{${}^2$}}
\label{eq:lenseq-sum}
\end{equation}
which, as we can see, is the lens equation (\ref{eq:lenseq}) for a sum
of Sun-like deflectors at distance $D$.  As throughout this section,
the source is much further away, a restriction we will lift later.
This verifies our ansatz (\ref{eq:refsdal-shapiro}) for the
gravitational time delay.

The combined geometric and gravitational time delay
\begin{equation} \label{eq:tau-prelim}
  t(\tang) = \frac D{2c} |\tang-\bang \,|^2
  - \frac G{c^3} \sum_i M_i \ln|\tang-\tang_i| \;,
\end{equation}
also variously known as the arrival time or the Fermat potential, is a
central concept in gravitational lensing.  As we have seen, the lens
equation is the zero-gradient condition on the time delay.  
The second and higher derivatives are also important, as detailed later.

Fermat's principle in gravitational lensing was introduced using
different approaches by \cite{bib1:Schneider1985} and
\cite{bib1:Blandford1986}.  Expressions like (\ref{eq:tau-prelim}) do,
however, appear in earlier work
\citep{bib1:Cooke1975,1983A&A...128..162B} albeit without the
interpretation as Fermat's principle.  Wavefront arguments were
introduced to gravitational lensing by \cite{1964MNRAS.128..295R}.
The gravitational time delay (\ref{eq:refsdal-shapiro}) is known as
the Shapiro delay after its deduction by \cite{1964PhRvL..13..789S} as
a property of null geodesics from general relativity.  That time
delays also emerge simply as an integral of the deflection angle was
shown independently (and actually slightly earlier) by
\cite{bib1:Refsdal1964}.

\section{The cosmological context}

\subsection{Why cosmological distances?}

The lensing galaxy in Q0957+561 is at redshift $z=0.36$, which is a
typical distance for strong lenses.  Why don't we see strong lenses nearby?

To see why, let us briefly consider the Sun.  The Sun is assuredly a
gravitational lens, but we do not observe it as a strong gravitational
lens, because the light deflection at its rim
(Eq.~\ref{eq:solaralpha}) is much smaller than its angular size on our
sky.  If one observed from a much greater distance $D$, the apparent
size of the Sun would become smaller than the deflection angle, and sources directly
behind the Sun would be lensed into images on either side.  We can
express this fact as
\begin{equation}\label{eq:threshold}
\frac{4GM}{c^2\xi} = \frac{\xi}D
\end{equation}
together with the condition that $\xi$ (the physical radial distance
of the apparent images on the observer's sky) is larger than the solar
radius.

What if the Sun were transparent, or we were looking for
strongly-lensed neutrinos?  Then the requirement that $\xi$ is more
than the solar radius would be lifted, and the condition
(\ref{eq:threshold}) would apply as
\begin{equation}\label{eq:crit_accel}
\frac{4GM(<\xi)}{\xi^2} = \frac{c^2}D
\end{equation}
with the mass enclosed within $\xi$. Note the latter form --- it says that
strong lensing can occur if the Newtonian acceleration at the lens is
at least $c^2/D$.  (This condition is usually expressed as a critical
2D density $M/\xi^2$, but let us stay with threshold acceleration for
now.)  From $D=\SI{1}{au}$ the threshold is too high even for the
solar interior, but for $D>\SI{25}{au}$ a transparent Sun will indeed
strongly lens neutrinos \citep{2000PhRvD..61h3001D}.  For
$D>\SI{550}{au}$ the optical solar gravity lens would be observable, and
is indeed the subject of some space-mission concepts
\citep{bib1:Turyshev2020}.  But for now, strong lensing needs much
larger distances and lower acceleration thresholds.

Let us consider some example cases of the threshold acceleration
(\ref{eq:crit_accel}).  The values\footnote{Here it is useful to
remember that $\SI{1}{au} \simeq \SI{500}{c\,\second}$ (light-seconds), and
$\SI{1}{pc} \simeq 10^8\,\si{c\,\second}$ (see Appendix~\ref{sec:app-cconv} for
more precise values).}
\begin{equation}
  M = M_\odot \qquad
  \xi \simeq \SI{2}{c\,\second} \qquad
  D \simeq 400\,{\rm au}
\end{equation}
correspond roughly to optical strong lensing by the Sun (the solar
gravity lens).  If we scale $\xi$ up by $10^3$ and $D$ up by $10^6$
we get
\begin{equation}
  M = M_\odot \qquad \xi \simeq \SI{4}{au} \qquad
  D \simeq \SI{2}{kpc}
\end{equation}
which is in the regime of microlensing in the Milky Way.  However, the
angular scale $\xi/D=\SI{1e-8}{\radian}\simeq\SI{2}{mas}$ which is
challenging to image even with optical interferometry
\citep[cf.][]{2022NatAs...6..121C}.  If, instead, we
scale all three quantities by $10^{12}$ we have
\begin{equation}
  M = 10^{12} M_\odot \qquad R \simeq \SI{20}{kpc} \qquad
  D \simeq \SI{2}{Gpc} \;,
\end{equation}
which is typical of massive galaxies at cosmological distances.
Moreover the angular scale $\xi/D=\SI{1e-5}{\radian}\simeq2''$ is
comparatively easy to resolve.

Thus, we see that strong lensing is very difficult to observe nearby,
but once you can observe galaxies at cosmological distance with
arcsecond resolution, strong lensing becomes more frequent.

\subsection{Distances in cosmology}

With lenses being at cosmological distances, we need to account for
the universe expanding as the light is travelling through it.  This is
quite straightforward to do, using the usual machinery of modern
cosmology, which is covered in many textbooks, and need not be
repeated here.  Let us nonetheless draw attention to the main concept
relevant to lensing, namely the different ways of expressing distance.

The observable associated with distance is the redshift $z$.  While an
observed $z$ may contain a small kinematic contribution (of order
$10^{-3}$), cosmological redshifts are essentially wavelengths
expanding with the universe.  Thus a redshift of $z$ refers to an
epoch when the scale factor of the universe was $1+z$ smaller than
now.  This applies to any cosmological model that is homogeneous and
isotropic on large scales (that is, satisfies the cosmological
principle).

In the current standard cosmological model, the universe is spatially
flat and its present total energy density is
\begin{equation} \label{eq:rho_crit}
  \rhocr = \frac{3H_0^2}{8\pi G} \;,
\end{equation}
in which $H_0$ is the current expansion rate, that is, the Hubble
constant\footnote{Distances, times, and angle-size relations needed
for studying lenses involve several unit conversions for $H_0$, which
sets the scales.  Appendix~\ref{sec:app-cconv} gives the relevant
conversions.}.  The density at redshift $z$ is $\rhocr$ times
\begin{equation}
  E(z) = \Omega_m(1+z)^3 + \Omega_r(1+z)^4 + \Omega_\Lambda \;,
\end{equation}
in which $\Omega_m$ is the current matter density fraction, $\Omega_r$ is
the current relativistic energy-density fraction, and $\Omega_\Lambda$
the ``dark energy'' density fraction.

The lookback time to some redshift $\zd$ is given by
\begin{equation}
  t_\mathrm{d} = \frac 1{H_0} \int_{\zd} \frac{dz}{(1+z)\,E(z)}
\end{equation}
and $ct_\mathrm{d}$ is by definition the distance travelled by light
from the source to the observer.  The comoving distance between two
points along the line of sight at redshifts $\zd$ and $\zs$ is given
by
\begin{equation}
  \mchi_{\mathrm{ds}} = \frac c{H_0} \int_{\zd}^{\zs} \frac{dz}{E(z)}
\end{equation}
As the name indicates, the comoving distance scales up distances
according to the present scale of the universe.

Gravitationally bound structures like galaxies and clusters do not
expand with the universe.  But the light coming from them
knows\footnote{This is known as anthropomorphism.} nothing of such
subtleties, and shows the angular sizes of galaxies and clusters as if
they \textit{had} been expanding with the universe.  To account for
this, angular-size or angular-diameter distances are defined.  The
angular-diameter distance to the lens (or deflector) and source are
\begin{equation}
   \dd = \frac{\mchi_\mathrm{d}}{1+\zd} \qquad
   \ds = \frac{\mchi_\mathrm{s}}{1+\zs} \;,
\end{equation}
respectively, with the subscripts having the obvious meanings.  Also
important is the angular-diameter distance
\begin{equation} \label{eq:dds}
  \dds = \frac{\mchi_\mathrm{s}-\mchi_\mathrm{d}}{1+\zs}
\end{equation}
from the lens to the source.  The latter needs clarification.
An observer at the lens would measure the angular distance as
$(1+\zd)/(1+\zs)$ times the comoving distance.  But that observer's
comoving distances are all smaller by $(1+\zd)$ than ours.  Hence the
expression (\ref{eq:dds}).

In the older literature \citep[notably in the well-known book
  by][]{bib1:Schneider1992} cosmologies with total density different
from $\rhocr$ are also considered, as are more general considerations
of the angular-diameter distances. The expressions given here
are restricted to cosmologies that are spatially flat (or equivalently,
have mean density equal to $\rhocr$), but this is a standard
assumption made in nearly all of the recent literature.

Yet another quantity which one occasionally meets in lensing
formalism is the conformal time.  A conformal time interval is given
by $(1+z)^{-1}\,dt$.  Using the conformal time, one can trace light
rays as if in a static universe.  In the previous section, we used
some arguments involving running light rays backwards, and the
arguments remain valid in an expanding universe if conformal time is
understood.

\subsection{The lens and source planes}\label{subsec:ds-planes}

With angular-diameter distances in hand, we can now extend the lens
equation and the time delay to the expanding universe, and relax the
assumption that sources are at infinity.

\begin{figure}
\includegraphics[width=0.48\textwidth]{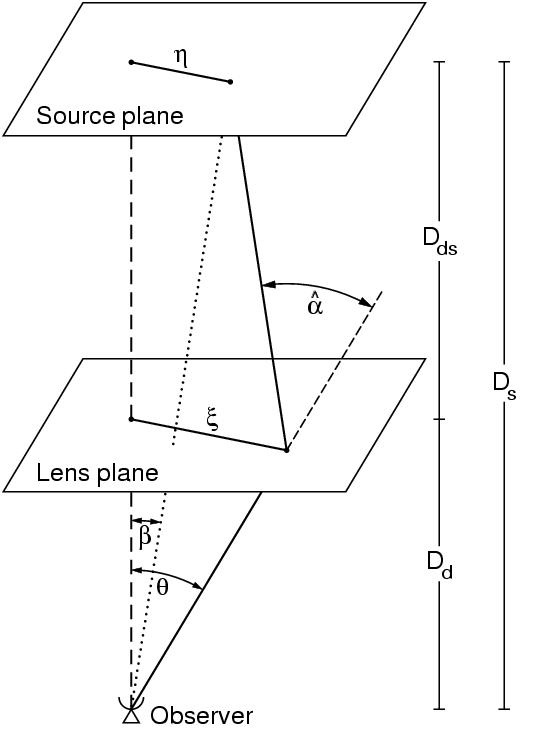}
\caption{A diagram \citep[reproduced from][]{2001PhR...340..291B}
  of the various lengths and angles used in the lensing
  formalism.  The dashed line is a reference direction, sometimes
  called the optical axis.  The angle $\hat\aang$ is the light
  deflection by the lens.  Note that $\dd,\dds,\ds$ are
  angular-diameter distances, and $\dd+\dds>\ds$, whereas $\xi$ and
  $\eta$ are physical distances.\label{fig:lens_geometry}}
\end{figure}

Fig.~\ref{fig:lens_geometry} shows the various quantities involved.
Through the small-angle approximation, the regions where the source
and lens are located, are approximated as planes.  The lens plane and source
plane have redshifts $\zd$ and $\zs$ respectively.  The lens
(deflector) is at angular-diameter distance $\dd$ from the observer,
so that an angle $\tang$ on the observer's sky corresponds to a
physical displacement $\xilen$ on the lens plane.  Analogously, the
source is at angular-diameter distance $\ds$ from the observer, and an
angle $\bang$ corresponds to a physical displacement $\etalen$ on the
source plane.  Note again that the angular-diameter distance $\dds$ is
not equal to $\ds-\dd$ but larger, because of the way angular-diameter
distances are defined.

Projecting distances on the source plane, we have
\begin{equation}
\ds\bang = \ds\tang - \dds \hat\aang(\tang)
  \label{eq:lenseqx0}
\end{equation}
and hence
\begin{equation}
  \bang = \tang - \frac{\dds}{\ds} \, \hat\aang(\tang)
  \label{eq:lenseqx}
\end{equation}
which is our lens equation (\ref{eq:lenseq}) generalized to an
expanding universe.  Introducing the apparent (or scaled) deflection
angle
\begin{equation}
\aang \equiv \frac{\dds}{\ds} \hat\aang
\end{equation}
makes the lens equation look exactly as before.

Assembling the lens out of Sun-like deflectors (as in
Eq.~\ref{eq:lenseq-sum}), we have
\begin{equation}
  \bang = \tang - \frac{4G}{c^2} \frac{\dds}{\dd\ds}
  \sum_i M_i \frac{\tang-\tang_i}{|\tang-\tang_i|\rlap{${}^2$}} \;,
\end{equation}
where we have used $\xilen=\dd\,\tang$.  As is easy to verify, this
equation is the zero-gradient condition of the time delay
\begin{equation} \label{eq:tau-intermed}
  \frac{t\,(\tang,\bang)}{1+\zd} =
  \frac{\dd\ds}{2c\dds} |\tang-\bang|^2
  - \frac{4G}{c^3} \sum_i M_i \ln|\tang-\tang_i| \;.
\end{equation}
We did not derive the $1+\zd$ factor in the time delay, but one can
argue that this factor must be there, because the gravitational time
delay occurs at the lens plane and gets redshifted.

One issue we have swept under the carpet here, is that, in
Eq.~(\ref{eq:lenseqx0}), we should have parallel transported the lens
plane to the source plane, instead of simply projecting.  When one
extends to multiple lens planes, as we will do below, disregarding
parallel transport produces spurious image rotation at leading order
\citep[cf.~][]{2018JCAP...07..067G}.  Such artifacts are harmless for
the applications considered here, but it is good to be aware of them.

\subsection{Multiple lens or source planes}

Although our equations so far have assumed one source plane and one
lens plane, it is not difficult to generalize to multiple planes.

The lensing of a source plane is independent of any other source
planes.  Hence, for a single lens plane, each source plane
$\bang_{\mathrm{s}}$ simply has its own lens equation
\begin{equation}
  \bang_{\mathrm{s}} = \tang -
  \frac{\dds}{\ds} \, \hat\aang_{\mathrm{d}}(\tang) \;.
  \label{eq:lenseq-sd}
\end{equation}
The form of Eq.~(\ref{eq:lenseq-sd}) tells us that a deflection at one
$\zd$ is equivalent to a scaled deflection at another $\zd$.  This is
useful in that minor lensing masses along the line of sight can be
replaced by effective masses on the main lens plane.

It is also possible to generalize the original lens equation
(\ref{eq:lenseqx}) explicitly to multiple planes.
To do so, let us first rewrite Eq.~(\ref{eq:lenseqx})
\begin{equation}
  \frac{\xilen}{\ds} = \frac{\etalen}{\dd} - \frac{\dds}{\ds} \hat\aang \;.
  \label{eq:lenseq-rewrite}
\end{equation}
Then let us introduce new subscripts: 0 for the observer plane, 1 for
the lens plane, and 2 for the source plane.  Let us further introduce
$\bx_0,\bx_1,\bx_2$ for comoving positions on these respective planes.
This gives
\begin{equation}
  \xilen = \frac{\bx_2-\bx_1}{1+z_2} \quad
  \etalen = \frac{\bx_1-\bx_0}{1+z_1} \;.
\end{equation} 
Meanwhile we can also write the angular-diameter distances in terms of
comoving distances
\begin{equation}
  \dd = \frac{\mchi_1-\mchi_0}{1+z_1} \quad
  \dds = \frac{\mchi_2-\mchi_1}{1+z_2} \quad
  \ds = \frac{\mchi_2-\mchi_0}{1+z_2} \;.
\end{equation}
Here $\bx_0$ and $\mchi_0$ are both zero, as they go from the observer
to the same observer, but they will play a useful role presently.

Substituting these various expressions into our rewritten lens
equation (\ref{eq:lenseq-rewrite}) and simplifying the result,
we obtain
\begin{equation}
  \frac{\bx_2-\bx_1}{\mchi_2-\mchi_1} =
  \frac{\bx_1-\bx_0}{\mchi_1-\mchi_0} - \hat\aang_1 \;.
\end{equation}
This equation relates comoving quantities at three consecutive planes.
Since the planes 0, 1, 2 were not special in the derivation, we can
generalize to
\begin{equation}
  \frac{\bx_{n+1}-\bx_n}{\mchi_{n+1}-\mchi_n} =
  \frac{\bx_n-\bx_{n-1}}{\mchi_n-\mchi_{n-1}} - \hat\aang_n \;.
\end{equation}
This is equivalent to Eq.~(19) from \cite{2014MNRAS.445.1954P}.  To
use it, we need to know the comoving distances $\mchi_n$ to all the
planes, the deflections $\hat\aang_n$ at those planes, and the
starting values $\bx_0=0$ (the observer) and $\bx_1$ (the scaled
location on the observer's sky).  We can then compute $\bx_n$ for
arbitrarily many lens planes.

So far, the best-known multi-plane system is the ``Jackpot lens''
J0946+1006 \citep{2008ApJ...677.1046G,2021MNRAS.505.2136S}, which has
three sources at different redshifts, one of which is also a second
lens.  Upcoming surveys will surely find more such systems, and the
topic is likely to become important.

\section{The standard formalism of gravitational lensing}

Several introductions to strong lensing have been published over the 
years, laying out the basic formalism, though the emphasis and amount 
of detail differs. 
\cite{bib1:Schneider1992} was the first book on lensing, and was very
influential.  The historical introduction is quite interesting, and
the quantitative portions of the book are very comprehensive and often
at an advanced level; however, the portions of the book that deal with
observational state of the field have been superseded many times over
in the last three decades.
\cite{2001PhR...340..291B} is an extension of the book to
weak lensing, but can also be used as an introduction to lensing
theory in its own right.
\cite{bib1:Blandford1986} give a short but well-explained article,
providing a gentle introduction among other things to catastrophe
theory as relevant to gravitational lensing \citep[][is a deep dive
  into that area]{bib1:Petters2001}.
They also discuss the similarities with atmospheric mirages, which is
further developed in \cite{bib1:Refsdal1994} who propose an original
introduction to the field and explain how to design an optical lens
simulator with a piece of glass.
\citep{2006glsw.conf...91K} provides an interesting perspective
specifically on strong gravitational lensing.
\cite{2021LNP...956.....M} is the most recent review, and includes
Jupyter notebooks. 
The above list is non-exhaustive; in addition to dedicated lensing reviews many text-books today contain chapters on lensing.

Common to all these sources, and indeed to the literature on strong
gravitational lensing generally, is a set of concepts and
cryptic-sounding terms that do not appear elsewhere in
astrophysics. With the physical basis and cosmological context in
hand, we now explain what these are.

\subsection{Introducing scaled quantities}

Let us replace the sum over discrete masses in
Eq.~(\ref{eq:tau-intermed}) with an integral over a projected mass
distribution
\begin{equation}
  \sum_i M_i \ln|\tang-\tang_i| \rightarrow
  \dd^2 \int \Sigma(\tang') \ln|\tang-\tang'| \, d^2\tang' \;.
\end{equation}
Here $\Sigma(\tang)$ is a function of the angular position, but
physically, it is a projected density in $\si{\kg\;\metre^{-2}}$ on the
lens plane. Hence the $\dd^2$ factor.

Next, we introduce two scales
\begin{equation} \label{eq:Tds-Sigc}
\begin{aligned}
  \Tds &\equiv (1+\zd)\frac{\dd\ds}{c\dds} \;, \\
  \Sigmacr &\equiv \frac{c^2}{4\pi G} \frac{\ds}{\dd\dds}
\end{aligned}
\end{equation}
and present their interpretations:
As defined, $c\Tds$ is a combination of lens and source
  distances, sometimes called the time-delay distance.  Like other
  cosmological distances, $\Tds\propto H_0^{-1}$.  In the limit
  $\dds\rightarrow\ds$, which is to say sources at infinity,
  $c\Tds\rightarrow\dd$.  Thus $c\Tds$ is a replacement for distance
  $D$ which we used in \S\ref{subsec:geomgrav} when introducing the
  general picture.

The quantity $\Sigmacr$ is known as the critical density for
  strong lensing, because it turns out to be the threshold for
  appearance of multiple images. (It is \textit{not} the cosmological critical
  density, which has different dimensions.)  For typical lens and source
  redshifts $\Sigmacr$ comes to a few $\si{\kilogram\;\metre^{-2}}$ --- about the surface density of
  window glass (see Appendix~\ref{sec:app-cconv}).  The proverbial astute reader will have noticed that
  $G\Sigmacr$ is an acceleration. Indeed the threshold acceleration we
  mentioned earlier (Eq.~\ref{eq:crit_accel}) is $4\pi G\Sigmacr$.

Using the scale parameters $\Tds$ and $\Sigmacr$ we can replace the
time delay and projected mass distribution with the dimensionless
quantities $\tau(\tang)$ and $\kappa(\tang)$
\begin{equation}\label{eq:Tds-Sigc-used}
\begin{aligned}
  t\,(\tang,\bang) &= \Tds \; \tau(\tang,\bang) \;, \\
  \Sigma(\tang) &= \Sigmacr \; \kappa(\tang) 
\end{aligned}
\end{equation}
and express the arrival time in the following, commonly used form
\begin{equation}\label{eq:taudef}
\begin{aligned}
  \tau(\tang,\bang) &= {\textstyle\frac12} |\tang-\bang|^2
  - \psi(\tang) \;, \\
  \psi(\tang) &= \frac1\pi \int \! \kappa(\tang')
  \ln|\tang-\tang'| \, d^2\tang' \;.
\end{aligned}
\end{equation}
Fermat's principle, $\nabla \tau(\tang) = 0$, gives the lens equation
(\ref{eq:lenseq}) in the form
\begin{equation}
  \bang = \tang - \nabla\psi(\tang) \;.
\end{equation}
The last term on the right is
\begin{equation} \label{eq:defl-integ}
  \aang = \frac1\pi \int 
  \frac{\tang-\tang'}{|\tang-\tang'|\rlap{${}^2$}} \;\;
  \kappa(\tang') \, d^2\tang' \;.
\end{equation}
Thus $\psi(\tang)$ behaves as a scaled version of a gravitational
potential.  It is known as the lens potential and is the solution of
the 2D Poisson equation
\begin{equation} \label{eq:poisson}
  \nabla^2 \psi(\tang) = 2\kappa(\tang) \;.
\end{equation}
Since the logarithmic kernel in Eq.~(\ref{eq:taudef}) is very broad, the
potential will be much smoother than the density.

The scaled density $\kappa(\tang)$ is known as the convergence.  We
will see the reason for this name in
\S\ref{subsec:mag}.  There is no single standard name for
$\tau(\tang,\bang)$, but Fermat potential, (scaled) time delay, or
arrival time are all used.

\begin{figure*}
\hbox to \hsize{\hss
\includegraphics[width=\hsize]{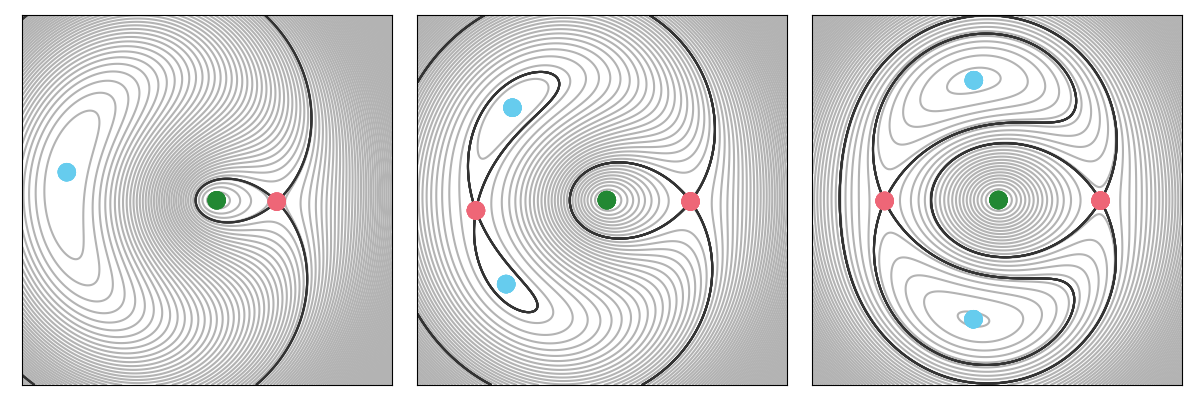}\hss}
\caption{Contour maps of arrival-time surfaces $\tau(\tang)$ image
  positions marked.  Minima are in light-blue, saddle points in
  orange-red, and maxima in green.  Contours through saddle points are
  highlighted.  The potential $\psi(\tang)$ is the same in all three
  panels, but the source positions are different.
\label{fig:arrivs}}
\end{figure*}

\subsection{The Einstein radius}\label{subsec:tae}

The quantities $\kappa,\psi,$ and $\aang$ behave analogously to the
density, classical gravitational potential, and force.  In particular,
a circular mass distribution behaves like a point mass when observed from outside.
This fact has useful consequences, and in particular it helps
understand the way $\kappa$ is defined.

Consider a point mass at the origin, that is
\begin{equation}
  \kappa(\tang) = \pi\tae^2 \, \delta(\tang) \;,
\end{equation}
in which $\tae$ is a constant whose meaning will be clear in a moment.
The lens equation for this mass is
\begin{equation}\label{eq:ptmass}
  \bang = \Big( 1 - \tae^2/|\tang|^2 \Big) \; \tang \;.
\end{equation}
A source at $\bang=0$ will result in a ring image at $|\tang|=\tae$
--- the well-known Einstein ring.  Its radius $\tae$ is known as the
Einstein radius.

Because of the aforementioned property of circular mass distributions,
the Einstein ring will be the same for any circular distribution of
mass within $\tae$ provided the integrated $\kappa$ equals $\pi\tae^2$
--- in particular a uniform distribution with $\kappa=1$ inside the disk.

This property lets us define a notional $\tae$ even when a mass
distribution is not circular: as the radius of a region whose mean
enclosed $\kappa$ is unity.

The mass $M$ corresponding to a given Einstein radius will be
$\pi\tae^2\,\Sigmacr$ and thus
\begin{equation} \label{eq:tae2M}
  \tae^2 = \frac{4GM}{c^2\dd} \frac{\dds}{\ds} \;.
\end{equation}
For a given mass, $\tae$ decreases with distance.  However, the
corresponding physical Einstein radius
\begin{equation}
   \re = \dd \tae
\end{equation}
increases with increasing distance.

One further interpretation of $\tae$ follows from rewriting
Eq.~(\ref{eq:tae2M}) as
\begin{equation} \label{eq:taeM}
  \tae = \frac{4GM}{c^2\xi} \frac{\dds}{\ds} \;.
\end{equation}
Recalling that $GM/\xi$ is the squared circular speed $v_c^2$ at
radius $\xi$ we see that $\tae$ is related to kinematics.  A circular
speed of $\SI{300}{\km\;\second^{-1}}$ corresponds to an Einstein
radius of up to $4v_c^2/c^2$ which is roughly an arcsecond.

\subsection{The three kinds of images}

Let us consider the Fermat potential or arrival time from
Eq.~\ref{eq:taudef}, that is
\begin{equation}
   \tau(\tang,\bang) = {\textstyle\frac12} |\tang|^2 - \psi(\tang) -
   \tang\cdot\bang
  \label{eq:tautilt}
\end{equation}
plus the constant ${\textstyle\frac12} |\bang|^2$, which can be
discarded.  It is interesting and useful to visualize
$\tau(\tang,\bang)$ as a surface: the time-delay or
arrival-time surface.

Without a lens, the arrival-time surface looks like a parabolic well
with a minimum at $\tang=\bang$. A lensing mass pushes the surface up
at nearby locations.  If the lensing mass profile is circular and directly in
front of the source, the minimum of the parabola will be replaced by a
hill with a circular valley around it.  However, unless the lens is
perfectly circular and the source is perfectly centered behind it, the
valley will not be perfectly symmetric, but will have a bowl on one side.
Elongated mass distributions can lead to two bowls, and for
more complicated mass distributions, further minima and maxima may appear.
Each bowl and hill corresponds to an image location, because, by
definition, $\nabla\tau(\tang)=0$ at minima and maxima.  In addition to
maxima and minima there are saddle points which also have
$\nabla\tau(\tang)=0$ and hence correspond to images.

The relationship between maxima, minima, and saddle points is
conveniently illustrated by contour maps of the arrival-time surface
$\tau(\tang)$,
such as in \figref{fig:arrivs}. 
The highlighted contours are self-crossing, and the
self-intersections are saddle points.  The saddle-point contours
separate regions of nested simple closed curves, and each such region
is either a bowl going down to a minimum, or a hill rising to a
maximum.  As we can see from the figure, there are two kinds of
saddle-point contours. 
One kind looks like an infinity sign, resembling a Bernoulli
  lemniscate (which is given in polar coordinates by
  $r^2=\cos(2\phi)$).  This kind occurs between two bowls, or between two
  hills.
The other kind of saddle-point contours resembles a lima\c con
  trisectrix ($r=1-2\cos\phi$ in polar coordinates), which corresponds
  to a hill within a bowl.

The lens in \figref{fig:arrivs} is an elliptical mass distribution (a
commonly-used form called a PIEMD --- see
Appendix~\ref{sec:app-exmods}).  Each panel shows $\tau(\tang,\bang)$
for a different $\bang$.  As is evident from the expression
(\ref{eq:tautilt}) we have a function of $\tang$ only plus a tilt due
to the $\tang\cdot\bang$ term.  In the right-hand panel of
\figref{fig:arrivs} we have a hill in the middle and two bowls above
and below it.  Turning to the middle panel we see the result of moving
the source a little to the left, which tilts the arrival-time surface
to the left; the two bowls move to the left.  The left panel shows how
moving the source further to the left tilts the surface more, and makes
the bowls merge into one.  A still larger tilt would make the hill
disappear too, leaving a single bowl.

More images can form, through nesting of the two cases
illustrated in \figref{fig:arrivs}.
\begin{enumerate}
\item A minimum is split by a lima\c con-like contour into a minimum and a
saddle point opposite, or
\item a minimum is split by a lemniscate-like contour into two minima
  with a saddle point in between.
\end{enumerate}
Less common is
\begin{enumerate}\setcounter{enumi}{2}
\item the inverted form of the second case, where a maximum is split
  into two maxima with a saddle point in between.
\end{enumerate}
The inverted form of the first case can occur in ordinary topography,
but does not occur in lensing.  A circular mass distribution produces
case~1; case~2 (whether nested or by itself) is characteristic of
elongated mass distributions.  For case~3 one needs multiple, very
distinct mass peaks.  Images always appear and disappear in
pairs: a saddle point and either a maximum or minimum. Thus, the number
of images is always odd.

\subsection{Magnification}\label{subsec:mag}

Although the number of images is always odd, the number of observable
images may not be.  This is because different images are variously
magnified, and some images can be demagnified so much that they cannot be detected anymore.
Magnification and demagnification are implicit in the
arrival-time surface, as we will see now.

The contour lines in \figref{fig:arrivs} hint at
magnification.  If one looks at this figure from afar (and tries to
ignore the heavy saddle-point contours as much as possible) the
regions where contour lines are close together look dark, while the
regions with widely-spaced contours appear to be bright.  Now, consider
a solution of the lens equation (\ref{eq:lenseq}) and take a
differential around it as
\begin{equation}\label{eq:delta-beta}
  \Delta\bang
  = \left( 1 - \nabla\nabla\psi(\tang) \right) \, \Delta\tang
  = \nabla\nabla\tau(\tang) \, \Delta\tang \;.
\end{equation}
If the Hessian $\nabla\nabla\tau(\tang)$ is large (small),
$\Delta\tang$ will cause us to move off-source quickly (slowly),
and meanwhile the contours will be close together (widely spaced).
Thus, the apparent brightness between the contour lines gives some
qualitative indication of magnification.

The Hessian
\begin{equation}
  \textbf{M}^{-1} \equiv \nabla\nabla\tau(\tang,\bang)
  = 1 - \nabla\nabla\psi(\tang)
\end{equation}
is an important quantity.  Formally inverting
Eq.~(\ref{eq:delta-beta}) we can write
\begin{equation}
  \Delta\tang  = \textbf{M}^{-1} \Delta\bang \;,
\end{equation}
to express the lensing result of small finite sources.
The inverse Hessian (or inverse curvature) of the arrival time is a tensor
magnification.  It is called the magnification matrix.

The inverse magnification matrix is conventionally separated into a
trace and a traceless part.  From the Poisson equation
(\ref{eq:poisson}) it follows that the trace will be $1-\kappa$.  The
traceless part is called the shear and denoted by $\gamma$.  Thus we
have
\begin{equation} \label{eq:maginv}
  \textbf{M}^{-1} =
  (1-\kappa)
  \begin{pmatrix} 1 & & 0 \\ 0 & & 1 \end{pmatrix}
  - \gamma
  \begin{pmatrix}
  \cos2\varphi &  \sin2\varphi \\
  \sin2\varphi & -\cos2\varphi
  \end{pmatrix} \;,
\end{equation}
with $\kappa,\gamma,\varphi$ all depending on $\tang$.
The meaning of the terms convergence and shear now
becomes evident: $\kappa$ represents an isotropic scaling, while
$\gamma$ represents a shearing oriented about position angle
$\varphi$.  The notation
\begin{equation}
  \begin{pmatrix}
  1 - \kappa - \gamma_1 &  \gamma_2 \\
  \gamma_2              &   1 - \kappa + \gamma_1
  \end{pmatrix}
\end{equation}
for $\textbf{M}^{-1}$ is also commonly used.  Either way, the matrix
can be diagonalized (taking
$\gamma_1\rightarrow\gamma,\gamma_2\rightarrow0$) by a suitable
rotation of the coordinate system.  The resulting diagonal elements,
which are the eigenvalues, are $1-\kappa\pm\gamma$.  For
minima both eigenvalues are positive; for maxima both are negative; at
saddle points the eigenvalues have opposite sign.  The determinant of
$\textbf{M}$
\begin{equation}
  \mu \equiv [(1-\kappa)^2 + \gamma^2]^{-1}
\end{equation}
is a scalar magnification. It is positive for minima and maxima, and
negative for saddle points, the absolute value representing the total
size change of a small source.  Clearly, if $\kappa$ or $\gamma$
become very large, the corresponding image will be demagnified and may
be unobservable.  At the centers of galaxies and clusters, $\kappa$
does become extremely large (though $\gamma$ remains modest).  This
produces an extreme peak in the arrival time, resulting in a
demagnified image (a maximum) at that location.  These demagnified
maxima are often observed at the centres of lensing clusters.  Lensing
galaxies, however, nearly always demagnify the central maxima to
unobservability.

A curious property of magnification is that there is always at least
one image with absolute magnification more than 1.  This appears to
say that lensing can increase the total amount of light.
However, this unphysical property is really an artifact of the
small-angle approximation \citep{2008MNRAS.386..230W}.

\subsection{Critical curves and caustics}

As we have seen, there are three kinds of images: minima, maxima, and
saddle points of the arrival-time surface $\tau(\tang)$.  As with
any topographic surface, the three kinds can be identified according
to the two eigenvalues of the curvature matrix
$\nabla\nabla\tau(\tang)$: minima have both eigenvalues positive;
maxima have both eigenvalues negative, whereas saddle points have one
of each sign.

Since $\tau(\tang)$ has only a quadratic dependence on
$\bang$, the curvature matrix $\nabla\nabla\tau(\tang)$ is
independent of $\bang$.  That is to say, although image location
depends on $\bang$, the image type ---if an image is present---
depends only on $\tang$.  It follows that image types must be
territorial: some regions can have only minima, and so on.  If there
are regions, there must be boundaries between the regions.  The
boundaries between regions of different image types are known as
critical curves.  One can find them by computing where the eigenvalues
of the curvature matrix change sign.  In general, both eigenvalues will not change sign together, so that minima and maxima do not border each
other and there is always a buffer of saddle points between them.

One can map critical curves to the source plane via the lens equation,
obtaining curves on the source plane, which are called caustics.
Critical curves are abstract curves on the sky, they are no
locations in real space.  Caustics, on the other hand, are real locations on the source plane.  
If a point source lies on a caustic, it will
light up a point on the critical curve.  There may be additional images at other
positions, but the image on the critical curve is special because
it has formally infinite magnification, since the curvature of the
arrival-time surface is zero.  This does not create a physical
paradox, because (a)~physically realistic sources have a finite size, and the singularity can
be integrated over, and (b)~wave optics would intervene to
remove the singularity.  If a moving source crosses from one side of a
caustic to the other, one of two things happens, depending on the
sense of crossing:  Either two images are created at the critical
curve and leave it in opposite directions, or the converse happens and two
images merge and disappear on the critical curve.
Positive-parity images always move in the same general direction as
the source, while saddle points move the opposite way, so there is no
ambiguity about which of these outcomes to expect.

\begin{figure*}
\hbox to \hsize{\hss
\includegraphics[width=\hsize]{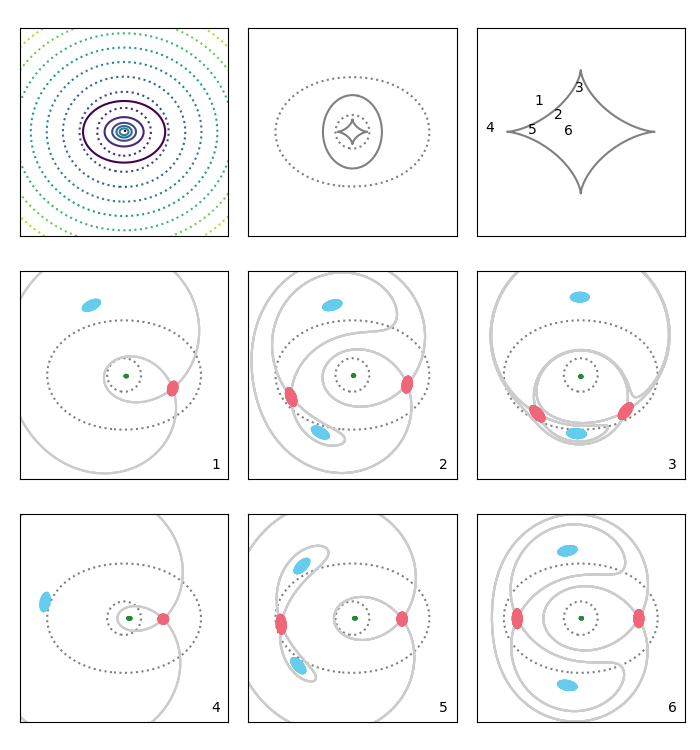}\hss}
\caption{Lens and image properties for a PIEMD.  The upper left panel
  shows contours of the surface density $\kappa(\tang)$ (solid curves)
  and the $\psi(\tang)$ (dotted curves).  The upper middle panel shows
  the critical curves (dotted) and caustic curves (solid).  The upper
  right panel is a detail of the `diamond' caustic with numbered source
  positions.  The numbers correspond to panels in the middle and lower
  rows.  The numbered panels show critical curves (dotted),
  saddle-point contours (solid) and images (as before, light blue:
  minima, orange red: saddle points, green: maxima).  The contrast
  between magnification and demagnification, and the stretching of the
  lensed images, is actually much more than appears here.
\label{fig:piemd}}
\end{figure*}

\begin{figure*}
\hbox to \hsize{\hss
\includegraphics[width=\hsize]{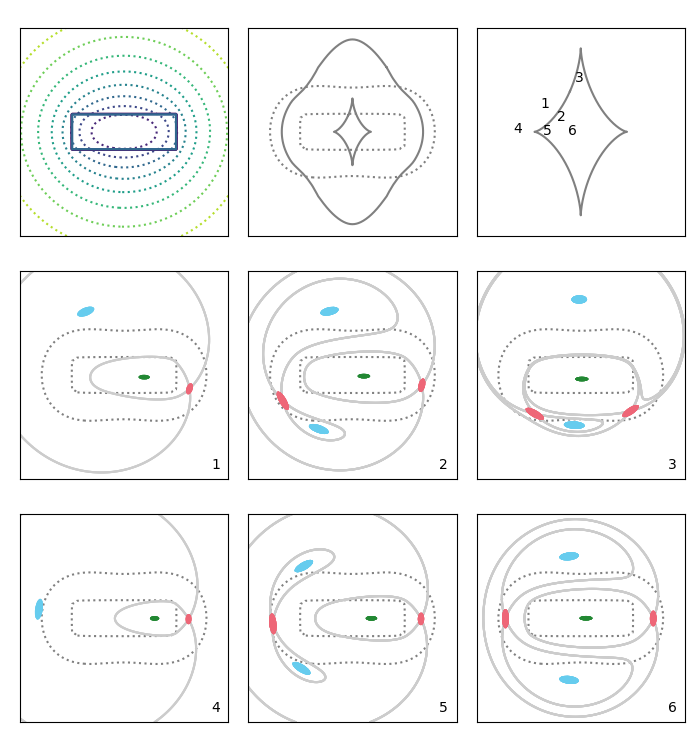}\hss}
\caption{As \figref{fig:piemd} but for a candybar lens.
\label{fig:candybar}}
\end{figure*}

\begin{figure*}
\hbox to \hsize{\hss
\includegraphics[width=\hsize]{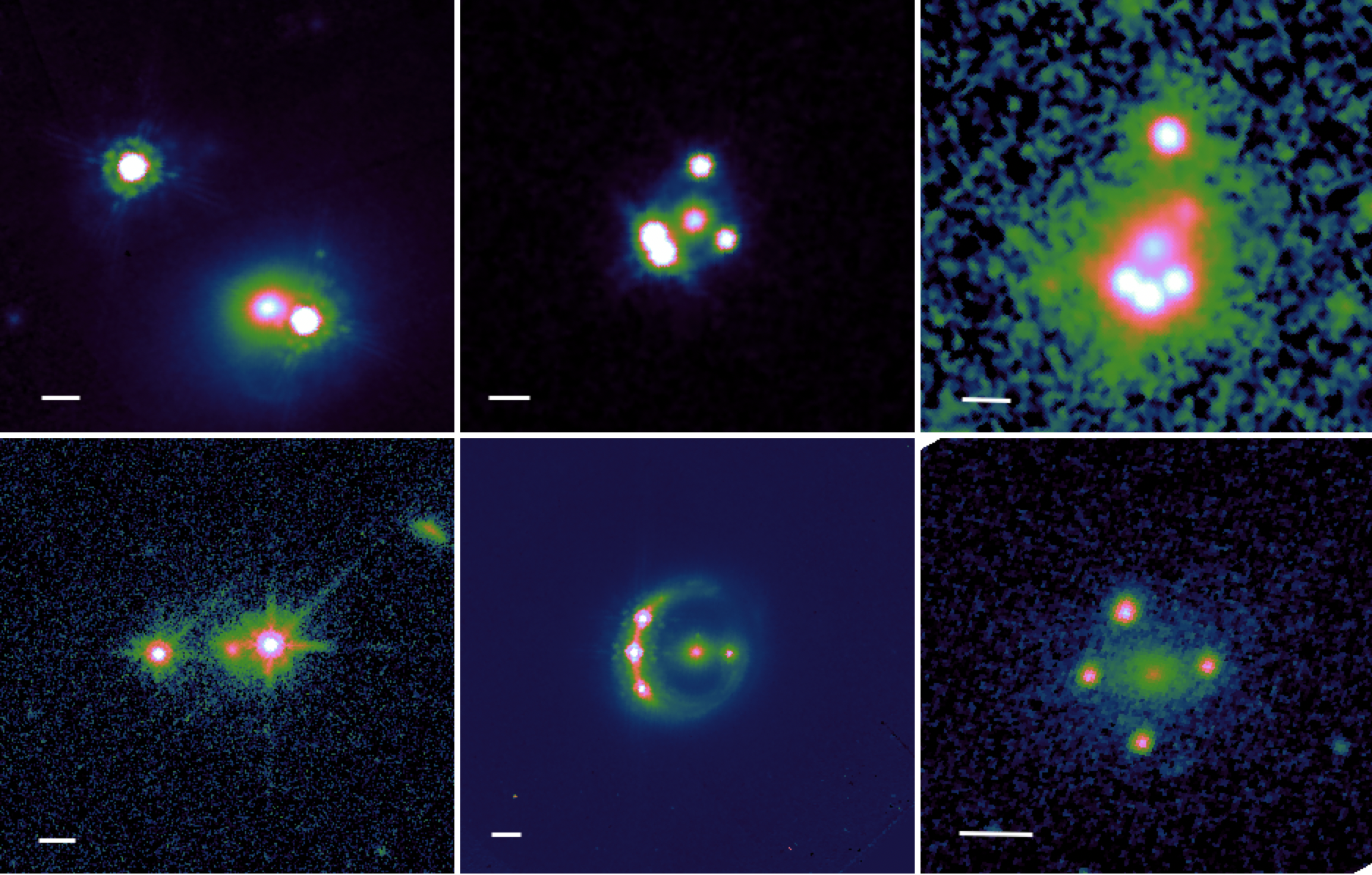}\hss}
\caption{A collage of HST images of lensed quasars oriented so that the
  lensed-image positions correspond to the example configurations in
  Figs.~\ref{fig:piemd} and \ref{fig:candybar}.  The systems (from top
  left in reading order) are Q0957+561, PG1115+080, RXJ0911+0551,
  HE1104-1805, RXJ1131-1231, and Q2237+030. The horizontal white tick marks $1^{\prime\prime}$ angular scale. Images were observed in wavelengths corresponding to $I-$, $J-$ or $H-$band filter.}
\label{fig:quasars}
\end{figure*}

In summary, critical curves are curves of extreme magnification on the
sky, whereas caustics are borders on the source plane at which the
number of lensed images changes.  In ordinary optics, caustics usually
refer to bright curves on a screen resulting from lensing of a small
bright source; this just means that source and observer are swapped
compared to our considerations here.  
In another difference of terminology, rainbows are sometimes referred to as caustics, but in the gravitational lensing convention they are critical curves.

The consequences of critical curves and caustics are enough to fill a
book \citep{2001stgl.book.....P} and still leave room for new
research.  Fortunately, the most common cases are fairly
straightforward, and in the following we will concentrate on those.

\subsection{Common image configurations}

The critical curves and caustics of a lens depend only on its
potential.  The images and saddle-point contours move as the source
moves.  
\figref{fig:piemd} revisits the lens used for \figref{fig:arrivs} and
illustrates example image configurations.  The first panel
shows the mass distribution and the corresponding lens potential, the
second panel shows the critical curves and caustics.  The rest of the
figure shows image positions, types, and magnifications for six
placements of the source.  The lower row has the same source
positions as \figref{fig:arrivs}. The outer critical curve separates
minima and saddle points, while the inner critical curve separates
saddle points and maxima.  The outer critical curve
corresponds to the inner caustic, and vice versa.  The inner caustic
has the qualitative form of an astroid ($x^{2/3}+y^{2/3}=1$ in
cartesian coordinates) but is also often called a diamond caustic.

A drastic modification is shown in
\figref{fig:candybar}. The elliptical PIEMD is
replaced with a rectangular tile shaped like a candy bar, with a
uniform $\kappa=4$ (for details see Appendix~\ref{sec:app-exmods}
again). The size is chosen so that the outer critical curve has the
same area as before.  We see that the potential contours are still
ovals, because of the logarithmic kernel in \eqref{eq:taudef}.  The
caustics are larger, and somewhat differently shaped.  The critical
curves are shaped very differently.  Yet, if we place the source at
similar locations with respect to the diamond caustic, the resulting
image configurations look similar.

\figref{fig:quasars} shows six lensed quasars with image
configurations similar to those in Figs.~\ref{fig:piemd} and
\ref{fig:candybar}.  These have been rotated (and some cases also
flipped) from the usual North-up/East-left orientation, for better
comparison with the panels labeled 1--6 in Figs.~\ref{fig:piemd} and
\ref{fig:candybar}.  Let us consider each of these examples in turn.
\begin{enumerate}
\item Q0957+561 is a typical double (not counting the presumed
  demagnified image coinciding with the lensing galaxy).  Doubles
  arise when the source is outside the diamond caustic.  One image
  lies outside both of the critical curves, while the other lies between
  the two critical curves.  The former is a minimum, while the latter
  is a saddle point in the arrival time.  The minimum will naturally
  have a shorter arrival time.  The two images form an obtuse angle
  with respect to the lensing galaxy, which happens when the source
  location is not aligned with the long or the short axes of the
  lensing mass.  We label it an ``inclined double''
  \citep[cf.][]{bib1:Saha2003}.
\item PG1115+080 is an ``inclined quad'', which is the type of
  configuration that arises if the source lies inside the
  diamond caustic, and is not near any of the cusps.  Compared to Q0957+561, two additional images,
  one a minimum and one a saddle point, form on either side of the
  outer critical curve.  This arrangement is sometimes called a
  ``fold type'' because the source lies near a fold caustic.
\item RXJ0911+0551, with three images close together near the lensing
  galaxy and one image further away on the opposite side, indicates a
  source near the cusp aligned with the short axis of the lensing
  mass.  It is a ``short-axis quad''.  Three nearly-merging images are
  characteristic of a source near a cusp.
\item HE1104-1805, with two images forming nearly a straight line with
  the lensing galaxy, is an ``axial double''.  This configuration
  arises when the source location is aligned with either the long or
  short axis of the lensing mass.
\item RXJ1131-1231 is a ``long-axis quad''.  It differs from a short-axis
  quad in that the three nearly-merging images are further from the
  lensing galaxy than the fourth image.  In this case, the source is
  near a cusp aligned with the long axis of the lensing mass.
\item Q2237+030 with its cross-like arrangement, is a ``core quad''
  and must have its source near the center of the diamond caustic.
\end{enumerate}

\begin{figure}
\hbox to \hsize{\hss
\includegraphics[width=.8\hsize]{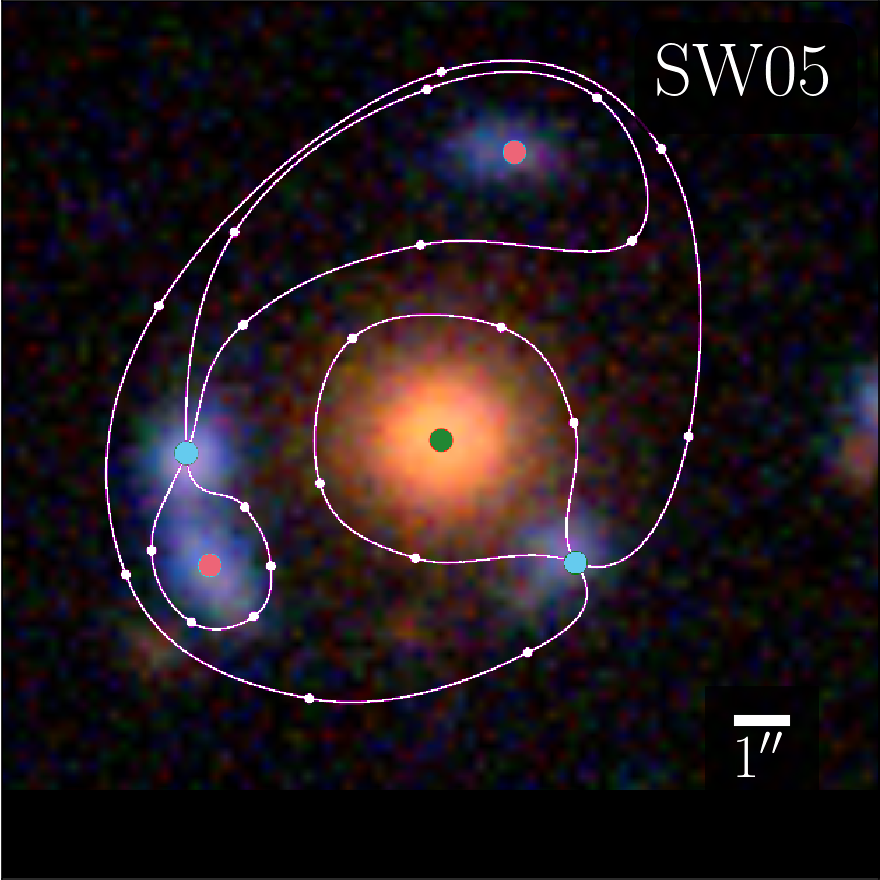}\hss}
\caption{The lens candidate SW05 J1434+5228 with a conjectural
  identification of the image structure \citep[taken
    from][]{2018MNRAS.474.3700K} which is similar to those on panel~2 of
  Figs.~\ref{fig:piemd} and \ref{fig:candybar}, as well as PG1115+080
  in Fig.~\ref{fig:quasars}.  This type of markup can be used as input
  for lens-modeling, as discussed later in \S\ref{subsec:ensem}.
\label{fig:spl-input}}
\end{figure}

With some practice, it is possible to stare at a lens candidate and
sketch conjectural saddle-point contours, thereby identifying the type
of each image (minimum, maximum, or saddle point) and predicting the
time ordering. \figref{fig:spl-input} shows an example. 

\subsection{Degeneracies}

The various examples in Figs.~\ref{fig:piemd}--\ref{fig:spl-input}
indicate two contrasting things about image configurations.  On the
one hand, some basic properties of the lensing mass, the general
placement of the source, and the likely time-ordering of the images
can be read off directly from the arrangement of images.  On the
other hand, the lensing mass distribution can be altered quite
drastically with little change in the images.  In fact, as we will see
presently, it is possible to modify a lens while keeping lensing
observables exactly fixed.  This is the problem of lensing degeneracies.

Let us go back to \figref{fig:wavefront}, and run the light backwards.
That is to say, let us consider a single observer above (at the origin
of the rays), and imagine sources scattered over the figure.  If the
sources are densely distributed and their locations are known, the
observer can expect to reconstruct the entire wavefront and infer the
gravitational field that caused it.  

The origin of lensing degeneracies is in that lensing gives
information about the gravitational field only along light paths to
the observer from lensed sources.
If the sources are sparse, the
observer's information will be incomplete.  Some clusters lenses go a
long way towards providing a dense sampling of sources over a range of
redshifts, and the situation will improve further with infrared
telescopes in space.  But for most lenses, one is working with a
single source plane, of which only a small part emits light.

A more formal treatment appears in \cite{bib1:Wagner2018d} and
\cite{bib1:Wagner2019b}.  The earlier paper characterizes the most
general class of degeneracies in the strong lensing formalism for a
single source and lens plane without making any assumption about the
mass distribution within the lens.  The second paper extends the
analysis from a single lens plane to additional lenses along the line
of sight, connecting the degeneracies in the lensing mass distribution
in a single lens plane to degeneracies in the overall matter
distribution along the entire line of sight \citep[see especially
  Fig.~3 in][]{bib1:Wagner2019b}.  These two works explain that one
physical cause of these degeneracies is rooted in the fact that
lensing only probes the integrated mass density along the entire line
of sight from the observer to the source.  Consequently, the
distribution of the not directly observable dark matter can be
redistributed in many ways.  At the same time, as shown in
\cite{bib1:Wagner2018c} and \cite{bib1:Wagner2019}, the observable
properties of multiply-imaged galaxies behind strong gravitational
lenses uniquely constrain \emph{local} properties of the lens as the
shear per mass density at the locations of the multiple images and
ratios of mass densities between the different multiple images. This
is the maximum information that can be extracted from static multiple
images without using any additional information about the overall mass
density distribution. All other properties of lenses, especially in
the areas far from any multiple image, are determined by the
assumptions contained in the lens models.

Among these general degeneracies, there are some simple but still
important ones, which we now discuss.

The first is a monopole degeneracy, which we have actually already
seen.  In \S\ref{subsec:tae} we noted that a circular mass
distribution behaves from outside like a point mass.  From inside it
has no deflection.  These two properties are analogous to the
well-known property of spherical shells in Newtonian gravity, and can
be derived using similar methods.
It follows that if we replace a mass ring with another mass ring, both
rings having the same center and the same mass but different radii,
there will be no change in the deflection, except in
the annular region between the two rings.
The same holds for derivatives of the deflection, including shear and magnification.
Thus, images not between the two
rings will be unaffected.  As a result, we are free to
choose any circular or annular region not
containing any images, and radially redistribute the mass in that
region, with zero effect on images.  Moreover, we can repeat this operation in
several round regions, as long as they
contain no images.  At distances outside all the images, we can freely
add or remove circular rings of mass.

Our next degeneracy is also very simple, though its consequences are
significant.  Consider the original lensing equation (\ref{eq:lenseq})
and multiply both sides by the same factor.  This will of course
preserve any solution of the equation, but will scale $t(\tang)$.
That is, the observed image positions will stay the same, but the time
delays between different images will get scaled, and so will the
implied source sizes.  The corresponding transformation of the lensing
mass is a little more subtle.  To derive it, we first recall
Eqs.~(\ref{eq:defl-integ}--\ref{eq:poisson}), and from them we first
derive
\begin{equation}
\nabla\cdot\aang = 2 \frac{\Sigma(\tang)}{\Sigmacr}
\end{equation}
We also have $\nabla\cdot\tang=2$.  Combining these
relations, it is clear that multiplying both
\begin{equation} \label{eq:mst}
\bang \ \hbox{and}\ 1 - \frac{\Sigma(\tang)}{\Sigmacr} 
\end{equation}
by a constant factor amounts to multiplying both sides of the lens
equation by the same factor.  This degeneracy goes by various names.
One is the mass-sheet degeneracy or mass-sheet transformation (or
MST), because its limiting case (multiplying the lens equation by
zero) makes $\Sigma(\tang)=\Sigmacr$ everywhere.  Another name is the
steepness degeneracy, since making the mass distribution less or more
like a uniform sheet makes the mass profile steeper or shallower.  Yet
another name is the magnification transformation, because rescaling
$\bang$ with $\tang$ fixed implies changing the magnification.

As expected from the wavefront picture, having more lensed sources
reduces the effect of lensing degeneracies.  Having more images reduces
the scope of the monopole degeneracy, because there is less area
available to redistribute mass.  The mass-sheet transformation
(\ref{eq:mst}) is prevented from being applied globally if sources at
different redshifts are present (because $\Sigmacr$ is not the same
for all images), but an extension that stitches together different
scalings in different regions of $\tang$ is possible
\citep{bib1:Liesenborgs2012,bib1:Schneider2014b}
There are also more complicated
redistributions of mass, coupled with transformations in $\bang$,
often constructed numerically in particular cases.  These are
collectively referred to as shape degeneracies \citep{bib1:Saha2006},
or source-position transformations \citep{bib1:Schneider2014}.

\section{Lens modeling}

\subsection{General considerations}\label{subsec:model-general}

Lens modeling is the process of reconstructing the lensing mass
distribution and the unlensed source brightness distribution from the
observables.  The principal observable is the lensed brightness
distribution.  Other observables --- time delays, or complementary
data like stellar kinematics --- may be available, depending on the
system.  Lens and source redshifts enter as parameters, and so does a
cosmological model with its parameters to convert these redshifts into
distances.

There are several different lens-modeling strategies in use and many
codes implementing them are available.  Some results, such as
measurements of enclosed mass, are robust across different modeling
strategies.  Other results may depend strongly on the
assumptions that go into the lens models.
For a quick visual impression, one can compare the mass maps and
magnification maps of the lensing cluster
Abell~2744\footnote{\url{https://archive.stsci.edu/prepds/frontier/lensmodels/}}
obtained by five different groups using the same observations.

There are many differences of details in how different modeling treat
the data.  But there is a deeper reason why different modeling methods
using exactly the same data can disagree about some of the results,
which applies even if the data are noiseless.  The problem is lensing
degeneracies, which we discussed earlier.
Given a gravitational lens and a light source,
computing the lensing observables is a well-posed problem. However,
the inverse problem of inferring the lensing mass and the unlensed
source does not generally have a unique solution, and additional
information has to be provided.

The issue is that we only have information about $t(\tang)$ where the
images are.  At these points, we know that $\nabla t(\tang)=0$ and the
values of $\nabla\nabla t(\tang)$.  We may also know the value of
$t(\tang)$ (up to a constant), if time delays are measured.  Roughly
speaking, time delays measure the potential difference between the
image locations, the image locations themselves are probes of the
gradient of the potential, while fluxes inform about the tidal field,
but these pieces of information are available only at image locations.
Elsewhere the information from lensing must be supplemented with other
data where available, and somehow interpolated where necessary.

In the following we will discuss the essential ideas behind different
modeling strategies in the literature.

\subsection{Spherical lens models}

Let us begin with the simplest.

Earlier, we have already briefly considered a point-mass less (see
\S\ref{subsec:tae} and Eq.~\ref{eq:ptmass}).  However, while a
point-mass lens is useful for modeling microlensing by individual
stars within our Galaxy, it is not an adequate approximation for strong
lensing by galaxies and clusters.  The light paths corresponding to
observed multiple images come through the dark halos of lensing
galaxies, or even through regions with significant stellar mass.  Thus,
the first requirement of a realistic model for a galaxy or cluster
lens is an extended mass distribution.

An elegant modification of a point mass, that does have an extended
mass distribution, is the Plummer model from stellar
dynamics,\footnote{This subsection quotes a number of results and
models from stellar dynamics, all of which are explained in more
detail in \cite{2008gady.book.....B}. } originally proposed a century
ago as a model for globular clusters.  The Newtonian gravitational
potential of a Plummer sphere is
\begin{equation} \label{eq:plummer-poten}
  \Phi(r) = \frac{GM}{(r^2+r_c^2)\rlap{${}^{1/2}$}} \qquad,
\end{equation}
with $r_c$ being interpreted as a core radius, which softens the singularity
of a point mass.  The corresponding density distribution is
\begin{equation}
  \rho(r) = \frac{3r_c^2}{4\pi} \frac{M}{(r^2+r_c^2)\rlap{${}^{5/2}$}} \qquad.
\end{equation}
Projecting on the sky from distance $\dd$ and writing the projected
$r$ as $\dd\theta$ and $r_c$ as $\dd\theta_c$ we have
\begin{equation}
  \Sigma(\theta) = \frac{M}{\pi\dd^2}
  \frac1{(\theta^2+\theta_c^2)\rlap{${}^2$}} \;.
\end{equation}
Dividing by the critical projected density gives
\begin{equation}
  \kappa(\theta) = \frac{\tae^2\,\theta_c^2}
        {(\theta^2+\theta_c^2)\rlap{${}^2$}}
  \qquad
  \tae^2 = \frac{4GM}{c^2\dd} \frac{\dds}{\ds} \;.
\end{equation}
The Einstein radius is the same as in Eq.~(\ref{eq:tae2M}) for a point
mass.  It is easy to verify that the mean $\kappa$ within a disk of
radius $\tae$ is unity.  The lens potential and deflection angle read
\begin{equation} \label{eq:plummerlens}
\begin{aligned}
  \psi(\theta)   &= {\textstyle\frac12} \tae^2 \, \ln(\theta^2 + \theta_c^2) \;, \\
  \aang(\theta)  &= \frac{\tae^2\,\tang}{\theta^2+\theta_c^2} \;.
\end{aligned}
\end{equation}
As expected, the $\theta_c\rightarrow0$ limit for all the expressions
is the point mass.

The Plummer lens is useful as a component in multi-component lens
models (which we will discuss below in \S\ref{subsec:multicomp}), but
by itself, it is not plausible as a galaxy halo.  Galaxy halos have
circular velocities which are roughly constant.  Now, the circular velocity is related
to the Newtonian potential by
\begin{equation}
  v_c^2 = r \frac{\partial\Phi}{\partial r}
\end{equation}
and from Eq.~(\ref{eq:plummer-poten}) we can see that $v_c(r)$ for a
Plummer sphere is far from being constant.

A better first approximation to galaxy halos is the isothermal sphere
and its variants.  
An isothermal sphere in stellar dynamics is a collisionless self-gravitating fluid of stars
with a Maxwellian velocity distribution.  The density is given by
\begin{equation}  \label{eq:sis3D}
G\rho(r) = \frac{\sigma^2}{2\pi r^2} \;,
\end{equation}
in which $\sigma$ is the stellar velocity dispersion which is constant
for a given system.  The circular
velocity is easily derived from the density (\ref{eq:sis3D}) and is
given by
\begin{equation} \label{eq:sis-vc}
  v_c^2 = 2\sigma^2
\end{equation}
for all $r$.  This is the ``flat rotation curve'' property.  The
constituent stars need not be on circular orbits. The velocity
distribution could be isotropic or even predominantly radial, but
\begin{equation}
  \langle v^2 \rangle = 2\sigma^2
\end{equation}
will apply in all cases due to the virial theorem.  The
observable stellar velocity dispersion will be the average along one
direction (the line of sight).  Thus
\begin{equation} \label{eq:sis-vlos}
   \langle v_{\rm los}^2 \rangle \simeq \textstyle{\frac23} \sigma^2 \;,
\end{equation}
with equality applying for an isotropic velocity distribution.

To derive the lensing properties of an isothermal sphere, we first
project the density (\ref{eq:sis3D}) in 3D to
\begin{equation}
G\Sigma(\theta) = \frac{\sigma^2}{2\dd\theta}
\end{equation}
on the sky and express it in terms of a new parameter $\tae$ as
\begin{equation} \label{eq:sis}
\kappa(\theta) = \frac{\tae}{2\theta} \;, \qquad
\tae = 4\pi\frac{\dds}{\ds} \frac{\sigma^2}{c^2} \;.
\end{equation}
As with the Plummer lens, it is easy to verify that the mean $\kappa$
within a disk of radius $\tae$ is unity. Hence $\tae$ really is the
Einstein radius.  The lens potential and deflection are
extraordinarily simple:
\begin{equation}
\begin{aligned}
  \psi(\theta)   &= \tae \, \theta \;, \\
  \aang(\theta)  &= \tae \, \frac{\tang}{\theta} \;.
\end{aligned}
\end{equation}
The deflection angle is simply $\tae$ times the radial unit
vector.  The deflection angle always having the same magnitude $\tae$ is the
lensing analog of the circular speed being the same everywhere.

The simple isothermal lens described so far is often called the
SIS, short for ``singular isothermal sphere''.  This refers to the density singularity at the
center.  The singularity can be removed by putting in a small core,
similar to the core in the Plummer sphere, yielding a non-singular
isothermal sphere, NSIS for short.  In models, however, the singularity is
often retained.  Redistributing the central density cusp to a circular
core has no effect on lensing behavior further out.  The only
practical effect of a central singular density is to make the
central image unobservable (cf.~\S\ref{subsec:mag} above).  Thus, if
no central image is observable, modelling the configuration with a central singularity is convenient.

Another singular property of the isothermal sphere is that the total mass is
formally infinite.  This follows from the density expression
(\ref{eq:sis3D}) and is the reason for $M$ not appearing in the
expressions for this model.  There are variants of the isothermal sphere
(called King models) which have a finite mass. This model class was
used in the very first lens models \citep{bib1:Young1980}.  It is,
however, more common to use the SIS as is, tacitly assuming that
the density is truncated outside the region of interest.

There are many more circularly symmetric lens models in use. We will not
discuss more examples here, as the Plummer and SIS models
illustrate the essentials.  The relation between kinematics and
lensing is especially worth noting.  We have seen earlier
(Eq.~\ref{eq:taeM}) that for a point lens, $\tae$ is proportional to
the orbital speed squared.  We now see that an expression like
(\ref{eq:sis}) relating $\tae$ and the velocity dispersion is a
consequence of the virial theorem.  A relation of this type can always be set up for
any lens in dynamical equilibrium, but the precise coefficient depends
on details of the mass distribution.

\subsection{Elliptical lens models}

If a lensing mass is circular, any and all images must be in the same
or opposite direction as $\bang$, because there is no other direction
in the system.  The arrival-time surface will be qualitatively like
the left panel of Fig.~\ref{fig:arrivs}.  There will be a minimum on
the same side as the source, a maximum near the lens center, and a
saddle point on the opposite side.  If the lens mass has a sharp peak
at the center (as in the SIS), the maximum will be at the center and
demagnified to vanishing brightness.
Observed multiple-image systems never have all images
(including demagnified images at the lens center) in a
straight line.  Thus, observed lenses are manifestly non-circular.

Generalising a circular lens to an elliptical one is easy, one only has to add a
term
\begin{equation}
\gamma_1 (\theta_x^2 - \theta_y^2) + 2\gamma_2 \theta_x\theta_y
\end{equation}
to the lens potential $\psi$ of a circular lens.  The additional term is called external shear (or XS).
By construction, external shear is traceless and hence implies no
change in $\kappa$, that is, in the mass distribution. Instead, it implies
mass \textit{outside} the region of interest.  External shear is the
lensing analog of a tidal field in orbital dynamics.

Additionally, one can modify the SIS or any other circular lens to
make it elliptical.  Thus, there is the SIEP (singular isothermal
elliptical potential) where the potential of the SIS has elliptical
isocontours.  Then there is the PIEMD (pseudo-isothermal elliptical
mass distribution) where the mass distribution $\kappa$ of the SIS is
both given a core and made elliptical, leading to a quite complicated
potential, given below in \S\ref{subsec:piemd}.  The PIEMD was used
for Figs.~\ref{fig:arrivs} and \ref{fig:piemd} earlier.  These lens
models are discussed in detail in \cite{bib1:Kassiola1993}
and \cite{1994A&A...284..285K}.
A large catalog of commonly-used mass models is given in
\cite{2001astro.ph..2341K}.

The disadvantage of ad~hoc modifications of the SIS is that they break
the connection with dynamics.  For circular lenses there are spherical
stellar-dynamical models, which make predictions for observable
kinematics in a lensing galaxy.  For the SIEP or the PIEMD there are
no known stellar dynamical models.  One then has to either build an
ellipsoidal galaxy model numerically, as done by
\cite{2007ApJ...666..726B} for example \citep[and more recently][has
  shown that an arbitrary elliptical lens can be expressed as a sum of
  stellar-dynamical components]{2019MNRAS.488.1387S} or fall back on
spherical models for the kinematics.

How well are real gravitational lenses approximated by elliptical lens
models?  To answer this question we turn to several studies, which
concentrate on quasar quads, because of the precise astrometry
provided by quasars.  Here there are two contrasting strategies that
have been developed: one is to model the observations with elliptical
vs more complicated lens models; the other other is to derive and use
some diagnostic for departures from
ellipticity, without necessarily fitting a model.
Let us consider the latter approach first.

One diagnostic for non-elliptic lenses
starts from the case of an SIEP, which has
\begin{equation}
\psi(\theta_r,\phi) \propto r (1 + q\cos(\phi-\phi_0)) \;.
\end{equation}
This potential exhibits a remarkable relation between the position
angles $\phi_i$ of the images about the lens center.  For this lens
\cite{1995MNRAS.272..363K} derive what they call a configuration
invariant, which can be written as
\begin{equation}\label{eq:config-invar}
  \cos(\phi_1+\phi_2-\phi_3) + \cos(\phi_2+\phi_3-\phi_1) +
  \cos(\phi_3+\phi_1-\phi_2) = 0
\end{equation}
taking the remaining position angle as $\phi_0=0$.  This expression
can be interpreted as a surface in the positional-angle
space $(\phi_1,\phi_2,\phi_3)$.
\cite{bib1:Woldesenbet2012,bib1:Woldesenbet2015} found that about half
of the quasars lie close to a polynomial surface, which they call
the fundamental surface of quads (FSQ), and image positions from the
SIEP satisfy it precisely.  \cite{2022AJ....164..120F} argue that the
FSQ must be an approximation to Eq.~(\ref{eq:config-invar}).

Another diagnostic
for deviations from elliptical symmetry comes from
\cite{bib1:Witt1996}, who
considered a general, purely elliptical potential
\begin{equation}
  \psi(\theta_x^2+\theta_y^2/q^2)
\end{equation}
and showed that all images, as well as the source, all lie on a
rectangular hyperbola.  It is not difficult to construct Witt's
hyperbola.  We start with a basic rectangular hyperbola
$\theta_x^2-\theta_y^2=1$, which has two
branches.  There are four simple linear transformations one can apply
to the curve: (i)~rotate, (ii)~shift along $\theta_x$, (iii)~shift
along $\theta_y$, and (iv)~rescale.  Together, they transform the
curve to
\begin{equation}
  \theta_x^2 - \theta_y^2 + c_1 \theta_x\theta_y +
  c_2 \theta_x + c_3 \theta_y = c_4 \;,
\end{equation}
which is a general rectangular hyperbola.  Writing down this equation
for any four points $(\theta_x^k,\theta_y^k)$ gives four linear
equations for the coefficients, and solving these equations gives a
rectangular hyberbola passing through all four points.  An important
predictive property of Witt's hyperbola is that it applies to any
central image as well, and since for a centrally peaked potential the
central image is at the lens center, the center of the lensing galaxy
will lie on the hyperbola.

\begin{figure}
\hbox to \hsize{\hss
\includegraphics[width=.8\hsize]{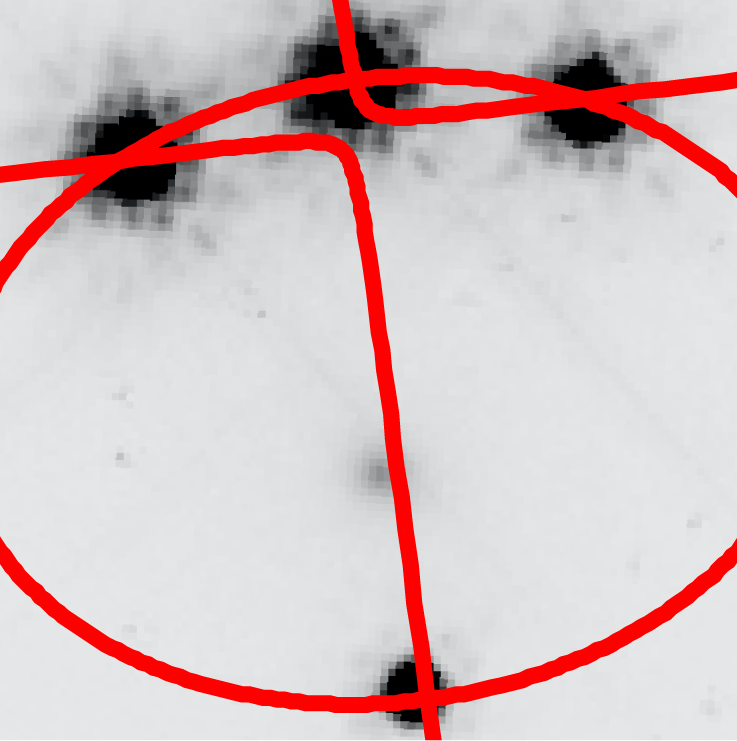}\hss}
\caption{The Witt-Wynne construction for the quad PS~J0147+4630
  \citep[from][]{bib1:Luhtaru2021}.  The image configuration is
  similar to those on panel~5 of Figs.~\ref{fig:piemd} and
  \ref{fig:candybar}.  The ellipse is similar though not identical to
  the critical-shear ellipse introduced by \cite{2002MNRAS.332..951W}
  and its short axis is aligned with the long axis of the potential.
\label{fig:witt-wynne}}
\end{figure}

Witt's hyperbola itself is independent of a specific choice of an
elliptical lens model, but can be used to find lens models via a
further construction \citep{bib1:Wynne2018}.  This is to construct an
ellipse (Wynne's ellipse) aligned with Witt's hyperbola that passes
through the four images of the quad.  The four images serve to set the
(i)~ellipticity, (ii)~shift along $\theta_x$, (iii)~shift along
$\theta_y$, and (iv)~scale of Wynne's ellipse.  The parameters of the
Witt-Wynne construction (see \figref{fig:witt-wynne} for an example)
translate into parameters for two distinct lens models: SIS+XS
\citep{bib1:Wynne2018} and SIEP with aligned XS
\citep{bib1:Luhtaru2021}.

The other approach, that of fitting elliptical and beyond-elliptical
models to the data (not only the lensed quasar positions but also
extended images) has been pursued by several authors.  An early
example is the study of PG~1115+080 by \cite{2005ApJ...626...51Y}, who
concluded that the lens is indeed an ellipse.
\cite{bib1:Sluse2012a} carried out a comparable study of
diverse quasar quads, while also identifying perturbers in the form of
nearby galaxies.  There are also several works \citep{bib1:Gomer2018,
  bib1:VandeVyvere2022a, bib1:VandeVyvere2022b} which introduce
non-elliptical features (twists, substructures) into simulated lenses,
to test how much they affect perturbed models and see how they affect
modeling.

While these studies are restricted to quadruply-imaged quasars, they
support a more general conclusion that simple elliptical lenses are a
good first-order approximation for most lenses, and the
counter-examples of large departures from elliptical lenses are
usually flagged by the presence of additional lensing galaxies that
perturb the main lens.  Accurately fitting observations thus requires
more features to be added.  \cite{bib1:Schmidt2023} and
\cite{bib1:Etherington2022} use elliptical lenses, including
components for additional lensing galaxies where necessary,
to model 30 and 59 lenses, respectively.  In contrast,
\cite{2018Sci...360.1342C} and \cite{2023arXiv230102656S} each study a
single galaxy lens in more detail, using 3D galaxy
models and spatially resolved kinematic data.  Meanwhile,
machine-learning approaches are also being developed
\citep[e.g.,][]{Gomer2023} towards more efficient
implementations of these models.

\def\blens{{\varpi}}
\def\bdata{{\cal D}}

\subsection{Bayesian terminology}

Lens models, especially if they are rather elaborate with many
parameters, are often formulated in Bayesian terms, the key words
being \textit{likelihood,} \textit{prior,} and \textit{posterior,} all
three being probability distributions.

To introduce these terms, let us write $\blens$ for the model
parameters and $\bdata$ for the data.  The $\blens$ are to be
understood in a very general sense, they can consist not only of
numbers but also of flags specifying the type of model, such as SIEP
or PIEMD and so on, and include the location and other properties of
the unlensed source.  The $\bdata$ consist of the images of the lensed system and of other available observations (e.g. time-delays).
The likelihood is
$p(\bdata\,|\,\blens)$, meaning the conditional probability of
$\bdata$ being as measured for given $\blens$.  Basically, the
likelihood is the probability of the noise difference between $\bdata$
and hypothetical noiseless data from $\blens$.  The prior is
$p(\blens)$, and it expresses our knowledge of the parameters before
having any data.  The posterior
\begin{equation}\label{eq:bayes}
  p(\blens\,|\,\bdata) \propto p(\bdata\,|\,\blens) \, p(\blens)
\end{equation}
is then given by Bayes' theorem in probability, and expresses our
knowledge of the parameters given the data.

The likelihood is usually conceptually clear, though it may involve a
lot of computation.  Typically it has a Gaussian form
$p(\bdata\,|\,\blens) \propto
\exp\left(-{\textstyle\frac12}\mchi^2\right)$.  The prior $p(\blens)$,
on the other hand, may be the subject of debate.  If the likelihood
factor in \eqref{eq:bayes} has a sharp peak in $\blens$, then the
posterior will have the same sharp peak, and the prior will have
little influence.  If, however, the likelihood does not have a single
sharp peak, the choice of prior will strongly influence the posterior.

Actually, the situation with $\blens$ is more complicated.
Because of lensing degeneracies, the likelihood can be insensitive to
some parameters.  Let us divide the parameters into `sharp' and
`blunt' $\blens = (\varpi_s,\varpi_b)$ such that the likelihood
$p(\bdata\,|\,\varpi_s,\varpi_b)$ is sharply peaked in $\varpi_s$ and
blunt in $\varpi_b$.  We can then marginalize out the blunt parameters
\begin{equation}
  p(\varpi_s\,|\,\bdata) = \int\! p(\blens\,|\,\bdata) \, d\varpi_b
\end{equation}
to extract the sharp (that is, well-constrained) parameters.  As a
simple example, consider a system with a clean Einstein ring and no
other data.  In this example $\varpi_s$ as the enclosed mass and
$\varpi_b$ as the radial profile are a possible parameter separation.
Suppose, however, that we use a different parametrization that mixes
sharp and blunt --- say we take $\blens = (\varpi_1,\varpi_2)$ where
$\varpi_1$ is a power-law slope and $\varpi_2$ is a normalization ---
and marginalize out the normalization.  We would then get an apparent
constraint on the power-law slope, that is really coming from the
prior.  In this simple example, the problem is easy to diagnose and
fix, but the situation gets worse with increasing the number of parameters
(distributed between $\varpi_s$ and $\varpi_b$), and hence the
hyper-volume of the parameter space.

\subsection{Source reconstruction}

When considering image formation earlier, we assumed that lensed images
are small enough to have an unambiguous position.  This is clearly
valid for lensed quasars and supernovae, and reasonable when speaking
of small components in a lensed galaxy, but it is not valid for arcs
formed from the lensing of entire galaxies.  In such cases one is
observing the superposition of a continuous family of lensed sources.
In fact, the recent modeling papers we mentioned above have to
reconstruct the sources, as well as modeling the lenses.

It is possible to reconstruct a lensed source independently of any
lens model, provided there is enough detail in the lensed images.
\cite{bib1:Tessore2017} and \cite{2021ApJ...919...38B} develop
formalism for doing so, but to explain the general idea, let us
consider an example, namely the spectacular five-image system in the
lensing cluster CL\thinspace 0024+1654.  \cite{bib1:Wagner2018c} draw
attention to six small components in the source galaxy, all of which
appear unambiguously in all five lensed images.  They then reconstruct
the source in a model-independent way, and compare the results with
those obtained by modeling the lens first.  Such systems are not
common, but nevertheless it is helpful to discuss why
model-independent source reconstruction works.  Let us think of the
six components on the source as being one reference $\bang_0$ and five
$\Delta\bang$ vectors.  These $\Delta\bang$ are magnified and sheared
into five sets of five observable $\Delta\tang$ vectors each by five
different magnification matrices.  The original $\Delta\bang$ vectors
are unknown ($5\times2$ numbers, because we are in 2D).  The five
magnification matrices are also unknown ($3\times5$ numbers, because
magnification matrices are symmetric and have 3 independent
components).  This yields a total of 25 unknowns.  Meanwhile, the
$\Delta\tang$ vectors provide $5\times5\times2=50$ constraints.  The
latter suffice to solve for all $\Delta\bang$ vectors, except for an
overall scale, which is degenerate.  The reference position $\bang_0$
of the source is not solved for, either, as it also degenerate.
Another example, this time for a galaxy lens, appears in the study of
B0128+437 by \cite{bib1:Wagner2020}, using four images with three
components each.

More commonly, however, the source and lens are reconstructed
together.  Consider an extended source with brightness distribution
$s(\bang)$.  When lensed, this source will produce an extended image
system
\begin{equation}\label{eq:exim}
  I(\tang) = \int \! \delta \big[\tang-\aang(\tang)-\bang\big] \;
  s(\bang) \; d^2\bang \;,
\end{equation}
where $\delta[\ldots]$ is the delta function.  The lens modeling
problem is now to recover $\aang(\tang)$ and $s(\bang)$ from
observations of $I(\tang)$ and any other available information.

Let us suppose that a lens model is available.  This may be a trial
lens model produced in one step of some larger iterative process.  But
let us say the lens is given, and in particular the deflection field
$\aang(\tang)$ is given.  Eq.~(\ref{eq:exim}) then takes the form
\begin{equation}\label{eq:exim2}
  I(\tang) = \int L(\tang,\bang) \, s(\bang) \, d^2\tang \;,
\end{equation}
where $L(\tang,\bang)$ is given.  In practice, there will be a further
convolution by a point spread function (psf) or other instrumental
response function, but if the psf is known, it can be absorbed inside
$L(\tang,\bang)$.  Under these assumptions, the source reconstruction
becomes a linear image-reconstruction problem.

For optical imaging, noise is generally Gaussian.  We can then write the
likelihood as $\exp\left(-\frac12\mchi^2\right)$, where
\begin{equation}\label{eq:src_chisq}
  \mchi^2 = \int \sigma^{-2}(\tang) \,
  \left( I^{\rm data}(\tang) - I(\tang) \right)^2 \, d^2\tang \;,
\end{equation}
with $\sigma(\tang)$ being the noise dispersion.  We are writing an
integral here for notational convenience, but it is actually a
sum over pixels.  There are several possible ways to go about
solving for $s(\aang)$.  One can assume a parametric form for
$s(\bang)$ and fit for the parameters by minimizing $\mchi^2$.
Alternatively, one can adopt a basis set for $s(\bang)$ and fit for the
coefficients in that basis set by minimizing $\mchi^2$.
\cite{2015ApJ...813..102B} present a suitable basis set for doing so
(shapelets), while \cite{2021A&A...647A.176G} consider wavelets.
A third option is to put $s(\bang)$ on a grid.  In this
approach, a simple minimization of $\mchi^2$ is actually not
recommended, because it tends to amplify the effect of noise, leading
to artifacts like negative-brightness pixels or checkerboard patterns.
To suppress such artifacts, a penalty function is added to $\mchi^2$
and the sum is minimized.  The choice of penalty function
--- or equivalently, the prior on the source light distribution ---
itself has a
large literature in image reconstruction \citep[for example, in
  interferometry --- see ][]{2017JOSAA..34..904T} and is known as
regularization.  \cite{2003ApJ...590..673W} in an influential paper
recommend
\begin{equation}\label{eq:smoother}
   \left|\nabla^2 s(\bang)\right|^2
\end{equation}
as the regularization function, which favors smoother $s(\bang)$.
It can be given a small or large weight, relative to $\mchi^2$, for
gentle or aggressive smoothing, as desired.
Further developments of the idea appear in several later works
\citep{2006MNRAS.371..983S,2009MNRAS.392..945V,2022MNRAS.516.1347V}.
An advantage of a quadratic regularization function like 
(\ref{eq:smoother}) is that the minimization procedure requires
solving only linear equations.  But it is not a requirement, and
source reconstruction using more complicated regularization functions
is also possible, notably in the early work by \cite{1989MNRAS.238...43K}
and later in \cite{2006ApJ...637..608B}.

\subsection{Many-component lens models}\label{subsec:multicomp}

Whereas for galaxy lenses a single elliptical component is often used,
while multiple components are becoming common, in cluster-lens models
multiple components are a must from the start.  This is partly because
clusters are further from a long-term stable state than galaxies and tend to have
more complicated mass distributions, but mainly because cluster lenses
present many more lensed images and thus more constraints.  An
illustrative example are the models of Abell~2744 by
\cite{2023ApJ...952...84B},
consisting of components for individual
galaxies, with a larger-scale component for the diffuse matter in the
cluster not identified with any individual galaxy.

The various modeling approaches we have discussed so far introduce
a lot of parameters to be fitted, but the system remains overdetermined,
meaning that there are more data points than parameters.  A different
category of multi-component methods reconstructs the lens in a
basis set such as a grid, where the number of components exceeds the
number of data points.  By itself, such a free-form reconstruction is
underdetermined and non-unique.  To solve that problem, regularization
can be applied, much as in the source reconstruction discussed above.
Using a regularization, one can search, for instance, the smoothest model that
fits the data, or the model the most similar to the light distribution.
An early example of this technique in lensing is the reconstruction of
Abell~370 by \cite{1998MNRAS.294..734A}.  The mass map of Abell~2744
by \cite{2018ApJ...859...58F} is typical of more recent work on these
lines.  Hybrids with basic elliptical components and an additional
substructure component which is free-form with regularization are
also possible \citep[e.g.][]{2022MNRAS.516.1347V,2022A&A...668A.155G}.

\subsection{Model ensembles}\label{subsec:ensem}

What if we specify neither a simple parametric form nor a
regularization?  In Bayesian terms, this means there are more
parameters in $\blens$ than there are data points in $\bdata$, while
$p(\blens)$ imposes some requirements on $\blens$ but has no hierarchy
of favored and less-favored models.  In practical terms, there is no
single optimal model, only an ensemble of models.  The ensemble
provides confidence regions on any desired quantity, which can be
easily extracted by marginalization.  Conceptually,
lens-model ensembles are like Monte-Carlo Markov chains widely used
for likelihood or posterior-probability sampling.

Model ensembles may
contain extreme models, but that does not mean the posterior
$p(\blens\,|\,\bdata)$ is uniform in all properties between the
extremes.  To give an analogy, consider an ensemble of random walks
with a given number and length of steps, and no preference between
different random walks.  The ensemble will span every random walk from
zero net displacement to a straight line (i.e., the maximum
displacement possible), but the distribution of displacements will not
be uniform between these extremes.

A very general implementation of these ideas for the reconstruction of
cluster lenses is the GRALE code \cite[introduced by][with many later
  enhancements]{2006MNRAS.367.1209L,2007MNRAS.380.1729L}.  In GRALE
each model consists of a large number of Plummer lenses
(Eq.~\ref{eq:plummerlens}) and there is an ensemble or population of
models that evolves according to a genetic algorithm to fit the
lensing data.  The prior is implicitly defined by the algorithm rather
than explicit, but is minimal: the mass distribution must be
non-negative.  The likelihood is not Gaussian, however it
does incorporate the data on image locations and shapes, and also the
absence of lensed images where they are not observed.  The minimal
prior means that the model ensemble tends to include a diversity of
models degenerate with respect to the observables.  Hence the inferred
uncertainties are generally larger than in other techniques.  The
reconstruction of Abell~2744 by \cite{2019MNRAS.488.3251S} is
illustrative.

Since galaxy lenses provide much less data than cluster lenses, a more
informative prior $p(\kappa)$ is necessary.  An explicit but still
very weak prior (amounting to requiring the lens to be centrally
concentrated) has been used in several modeling codes
\citep{2000AJ....119..439W,bib1:Saha2004,
  2014MNRAS.445.2181C,2018A&C....23..115K}.  Here we mention two
contrasting applications. \cite{bib1:Denzel2021} used the
model-ensemble method to model eight time-delay quasars simultaneous,
with $H_0$ as a shared parameter to be inferred. In
\citep{2018MNRAS.474.3700K} citizen scientists modeled tens of galaxy
lenses by marking up each lens candidate and using a model-ensemble
code to quickly explore the range of models consistent with the
markup.  Fig.~\ref{fig:spl-input} shows an example of input markup
from that work.

Finally, we mention yet another recent strategy, which amounts to finding plausible matches in galaxy-formation simulations instead of trying to reconstruct the lens. 
That is, the prior
$p(\kappa)$ can be a large ensemble of simulated galaxies, which is
then filtered to a much smaller ensemble that is consistent with the
available data.  Any such technique will be computationally expensive,
perhaps unrealistically so for clusters, but is starting to become
feasible for galaxy lenses \citep{2021MNRAS.506.1815D}.  What is
attractive is the prospect of bypassing lens modeling altogether, and
using lensing observations to directly address questions about galaxy
structure and formation.

\subsection{Model-comparison studies}

It is acceptable for different lens models (i.e. models that differ in
lens/source reconstruction) to disagree on some derived quantities as
long as the quantities important for a specific science goal are
accurately recovered. Strong lensing very accurately recover the
Einstein radius and the mass enclosed within that radius, but other
quantities (e.g. the ellipticity of the mass distribution) may be less
accurately recovered. Some of these uncertainties on recovered
parameters arise because of the degeneracies mentioned earlier, this
is what is commonly referred to as ``known unknowns'', but other
differences arise because of ``unknown unknowns''. The latter are
harder to identify and require model comparisons.

We have already mentioned Abell~2744 several times, in connection with
different modeling approaches.  The system is one of six lensing
clusters (the Frontier Field clusters) which have been likewise
modeled by several groups.  There have further been what one may call
meta-analyses \citep{bib1:pri17,rem18,bib1:Raney2020} which study and
interpret the effect of different choices made by different
methods. These comparisons demonstrate a global robustness of models
in regions well constrained by the data. They also highlight the
complementarity of the parametric and free-form models.

Another kind of model-comparison study are lens modeling challenges,
where different groups are invited to reconstruct a lens from mock
data.  \cite{bib1:Meneghetti2017} and \citep{bib1:Ding2021} are two
such blind tests of the methods of several groups, covering cluster
lenses and galaxy lensed respectively.  Such studies are a useful
indicator of robustness and uncertainties in the current state of the art.

A third kind of model comparison consists in blindly forecasting some
observables with various models. One spectacular case of this exercise
happened with the observations of the lensed supernova
``SN~Refsdal''. After the observations of the first four lensed images
of the supernova, several groups gathered to predict the reappearance
position, brightness of the 5th image (located $\sim8''$ from the
first quadruplet) about hundreds of days after the previous image
\citep{Kelly2023ApJ, Kelly2023Sci}. A fraction of the models
successfully reproduced the data, enabling the use of this first
multiply imaged lensed supernova for $H_0$ inference. The success of
such a model comparison is very encouraging as numerous supernovae may
be discovered in upcoming ``all-sky'' surveys (see the chapter on
transients).

The comparison of models is a useful but difficult exercise because
differences between models may arise not only from fundamental
degeneracies and prior choices as discussed above, but also from
subtle bugs that do not teach much about lensing but reveal a need to
fix a code. Importantly, it is also hampered by the difficulty of
comparing codes on the same grounds. The comparison of deflection and
magnification maps offers a simple mean of comparison but extra work
is required to derive differences and biases on specific parameters,
such as e.g. the slope of the mean slope of the density profile, or
the position angle of the main axis of the mass distribution. An
additional hurdle is the difference of conventions in defining some of
these quantities. The COOLEST framework \citep{Galan2023}, which
provides an infrastructure to gather and compare results of lens
models in a painless way, may help and trigger more model comparisons
on large ensembles of lenses and ensure a good control of systematic
uncertainties.

\subsection{Reflections on lens modeling}

As we remarked earlier (in \S\ref{subsec:model-general})
strong-lensing observations measure the projected gravitational
potential its derivatives at the image locations, and the task of lens
modeling is, in effect, to interpolate the values elsewhere.
Doing so requires additional assumptions, and preferably complementary
data as well.

A useful analogy is the density inside the Earth.  We can measure
gravitational acceleration at the surface or from orbit, and this
gives the total mass very accurately, and also provides information on
multipoles, but it does not give the radial density profile.  That
requires theory on the structure of planets, and further data like
seismography.

Likewise lens modeling depends on our knowledge of and uncertainties
about the structure and dynamics of galaxies and clusters.  Are
lensing galaxies all isothermal ellipsoids to the few-percent level?
Perhaps not.  Are dark halos so full of ongoing mergers that all
equilibrium models are useless?  Perhaps not, either.  But between these
extremes, there is a diversity of views among researchers in lensing.

The plurality of methods and data used for modeling lens systems may look difficult to unite at first sight. However, the reconstruction of the lensing mass density generally results from a trade-off between model rigidity and flexibility chosen in the light of the data and scientific questions posed, a situation that machine learning scientists may frame in terms of bias and variance arising for under-/over-fitting of the data. But beware, the choice of a model over another is not uniquely governed by those considerations. Inductive biases, namely biases introduced by the choice of a method, also exist. They are inherent to scientific research and evolve hands in hand with scientific results. A diversity of views and methods is often a good sign that our understanding of the world progresses. Strong lensing is no exception to this research walk.

\section{Conclusions and outlook}

The introduction to strong gravitational lensing presented here starts
by inviting the reader to think how gravitational fields will affect
geometrical wavefronts propagating through the Universe.  This leads
to the main theoretical construct, the Fermat potential or
arrival-time surface.  The first and second derivatives of the arrival
time surface are the lens equation and the distortion (or
magnification) matrix, respectively, which are the main players in the
quantitative analysis of the problem.  The equations are the same as
in other approaches that use only geometrical optics to describe light propagation, but many
properties of lensing follow from geometrical arguments even without
equations.  In particular, visualising the arrival-time surface for a
specific lensing configuration allows the researcher to gain a
qualitative understanding of the problem.

The paper also introduced the reader to the other two essential components 
of strong lensing: critical curves and caustics. With these theoretical 
concepts under their belt, the reader is ready to learn about lensing 
as a probe in astrophysics. As its fundamentals are well understood, the 
primary application of lensing in astrophysics is to build mass 
models of the lens, and to recover the structure of the background source. Most of the 
mass in galaxy and cluster lenses is dark, so lensing is very handy here. 
Distant, high-redshift sources tend to be faint and small; cosmic lenses make
sources larger, allowing for better spatial resolution. We also consider 
some simple lens models, and briefly describe more sophisticated ones 
used in the literature.

This short introduction to the essentials of strong lensing should provide 
the reader with the basic principles and physical intuition as a starting 
point to dive deeper into this growing field. The applications of lensing 
are ubiquitous: hardly any aspect of modern astrophysics has been untouched by it. 
The other papers in this series cover most, if not all the current, as well 
as emerging applications of lensing: 
Search for strong gravitational lensing through serendipitous and deliberate discoveries,
Strong lensing by galaxies,
Strong lensing by galaxy clusters,
Strong lensing and microlensing of quasars,
Microcaustics near macrocaustics,
Gravitational lensing as a cosmic telescope, 
Probing the nature of dark matter on (sub-)galactic scales with strong lensing, 
Time-Delay Cosmography and the estimation of the Hubble Constant and other cosmological parameters,
Strong lensing and microlensing of supernovae, 
Strong lensing of Gamma-Ray Bursts, Fast Radio Bursts, and Gravitational waves,
and
Diffraction in strong gravitational lensing.

We hope the reader enjoys this and other papers in this series, and finds 
strong gravitational lensing as enthralling as the authors do!

\vspace{0.35cm}
\noindent{\bf{Acknowledgments}}
We would like to thank the International Space Science Institute in Bern, Switzerland (ISSI) for their hospitality, and the conveners for organizing the stimulating workshop on ``Strong Gravitational Lensing''.
We thank our co-participants at the workshop for many discussions and
suggestions, and the reviewers for their comments.
We especially thank P.~L.~Schechter and P.~Schneider for sharing many insights.

\vspace{0.35cm}
\noindent{\bf{Competing Interests}}
The authors declare no competing interests.

\bibliographystyle{aasjournal}
\bibliography{main.bbl}                

\appendix

\section{Example lens models used}\label{sec:app-exmods}

\subsection{A softened Schwarzschild metric}

Consider the metric
\begin{equation}
  ds^2 = -e^{2\Phi(r)} dt^2 + e^{-2\Phi(r)} dr^2 +
  r^2 \left( d\theta^2 + \sin^2\theta\,d\phi^2 \right)
\end{equation}
If we put
\begin{equation}
  2\Phi(r) = \ln \left(1 - \frac{2GM}{c^2r}\right)
\end{equation}
we would have a Schwarzschild metric.  For the example in
Figs.~\ref{fig:rays}--\ref{fig:twowf} we instead we put
\begin{equation}
  \Phi(r) = - \frac{GM}{c^2}\frac1{\sqrt{r_c^2+r^2}}
\end{equation}
where $r_c$ is a small softening length.

This metric has no especially interesting physical interpretation.  It
is just used as an example of a strong gravitational field without the
additional complication of an event horizon.

\subsection{PIEMD} \label{subsec:piemd}

\def\re{R_e}
\def\rf{R_f}
\def\sq{\sqrt{1-q^2}}
\def\arctanh{\mathop{\rm arctanh}\nolimits}

The so-called pseudo-isothermal elliptical mass distribution, used as
an example in Figs.~\ref{fig:arrivs} and \ref{fig:piemd} is defined
via two elliptical radii
\begin{equation}
\begin{aligned}
  \re^2 &= q^2(s^2 + \theta_x^2) + \theta_y^2 \\
  \rf^2 &= (1-q^2)\theta_x^2 + (\re+s)^2
\end{aligned}
\end{equation}
Then the density is
\begin{equation}
\kappa = \frac1{2\re}
\end{equation}
the deflection is
\begin{equation}
\begin{aligned}
  \alpha_x &= \frac1{\sq}\arctan\left(\frac{\sq \theta_x}{\re+s}\right) \\
  \alpha_y &= \frac1{\sq}\arctanh\left(\frac{\sq \theta_y}{\re+q^2s}\right)
\end{aligned}
\end{equation}
while the potential and its second derivatives are
\begin{equation}
\begin{aligned}
   \psi &= \theta_x\alpha_x + \theta_y\alpha_y - s\ln\rf \\
   \psi_{xx} &= \frac{(qs)^2 + \re s + \theta_y^2}{\re\rf} \\
   \psi_{yy} &= \frac{s^2 + \re s + \theta_x^2}{\re\rf} \\
   \psi_{xy} &= -\frac{\theta_x\theta_y}{\re\rf}
\end{aligned}
\end{equation}

\subsection{A candy-bar lens}

The candy-bar lens, used as an example in \figref{fig:candybar}, is
not by itself a serious lens model, but can be a useful component of
one.  Square mass tiles are used in many free-form models.  Another
application is in microlensing, where a rectangular mass mass tile is
removed from a smooth mass distribution and replaced with point
masses.  To compute the potential, we need the integral from
Eq.~\ref{eq:taudef} for constant $\kappa$, that is to say
\begin{equation}
\int \ln\left(\theta_x^2 + \theta_y^2\right) \, d\theta_x \, d\theta_y
\end{equation}
over a rectangular region.  The result and its derivatives can be
expressed in terms of arctans and logs.  Expressions for a square are
given \cite{1997MNRAS.292..148S} and for a rectangle in
\cite{2022ApJ...931..114Z}.

\section{Constants and unit conversions}\label{sec:app-cconv}

Astronomical distances are commonly given in many different units.
When contrasting scales are involved, as they often are in
gravitational lensing, it is rather convenient to convert everything
to light-seconds.  Thus, while a parsec is defined as the distance
from which an astronomical unit (au) subtends an angle of
$1\rm\,arcsec$, with the au being $\SI{1.495978707e11}{\metre}$, the
values
\begin{equation}
\begin{aligned}
  \SI{1}{\rm au} &= c\times\SI{499.0}{\second} \\
  \SI{1}{\rm pc} &= c\times\SI{1.029e8}{\second}
\end{aligned}
\end{equation}
are easier to work with, and more digits can be added if necessary.
Similarly, the solar mass can be expressed as
\begin{equation}
  \frac{GM_\odot}{c^3} = \SI{4.925e-6}{\second}
\end{equation}

The Hubble constant is commonly given in the form
\begin{equation}
H_0 = h_{70} \times \SI{70}{\kilo\metre\;\second^{-1}Mpc^{-1}}
\end{equation}
The following conversions may be useful.
\begin{equation}
\begin{aligned}
  H_0^{-1} &= h_{70}^{-1} \times \SI{14.0}{Gyr} \\
  cH_0^{-1} &= h_{70}^{-1} \times \SI{4.28}{Gpc} \\
            &= h_{70}^{-1} \times \SI{20.8}{kpc\;arcsec^{-1}}
\end{aligned}
\end{equation}
The critical density (from Eq.~\ref{eq:Tds-Sigc}) comes to
\begin{equation}
\frac{cH_0}{4\pi G} = h_{70} \times \SI{0.811}{\kilogram\;\metre^{-2}}
\end{equation}
times a dimensionless factor which is typically $\sim1/\zd$.

The cosmological total density $\rhocr$ (from Eq.~\ref{eq:rho_crit})
can be expressed in several ways, such as
\begin{equation}
\frac{3H_0^2}{8\pi G}\frac{c^2}e =
h_{70}^2 \times \SI{5.15}{GeV\;\metre^{-3}}
\end{equation}
\section{Lensed Waves}\label{sec:app-waves}

If a source is large enough, that the distance to the observer from
different parts of the source differ by more than a wavelength,
radiation from it is incoherent at the observer.  This is the case for
nearly all astronomical sources.  For these, geometrical optics
applies.  In particular, the arrival-time surface, whether in its
scaled form $\tau(\tang,\bang)$ or in its dimensional form
$t(\tang,\bang)$ (see Eqs.~\ref{eq:Tds-Sigc}--\ref{eq:taudef}), is
defined over the whole sky, but only its zero-gradient points
correspond to images.

For sources that are small enough and far enough, however, the whole
$t(\tang)$ surface interferes.  This applies to gravitational waves
\citep[see e.g.,][]{2020arXiv200712709D} and may apply to fast radio
bursts \citep{2022PhRvD.106d3017L}.  At frequency $\omega$ the
resulting electric field of the light (or gravitational wave strain)
is given by
\begin{equation}
  \int \exp\left(i\omega\,t(\tang)\right) \, d^2\tang
\end{equation}
This is sometimes called the Fresnel-Kirchoff diffraction integral.
Of particular interest is the quotient
\begin{equation}\label{eq:mag-quotient}
  \frac{\int \exp\left(i\omega\,t_L(\tang)\right) \, d^2\tang}
       {\int \exp\left(i\omega\,t_{NL}(\tang)\right) \, d^2\tang}
\end{equation}
of the lens and no-lens cases.

With no lensing
\begin{equation}
  t_{NL}(\tang) = {\textstyle\frac12}\Tds
  \left( (\theta_x-\beta_x)^2 + (\theta_y-\beta_y)^2 \right)
\end{equation}
with the scale $\Tds$ being as defined in from
Eq.~(\ref{eq:Tds-Sigc}).  With a lens present, things are of course
more complicated.  For one regime, however, namely where the wave
period is much shorter than gravitational time delays (that is,
$\omega t(\tang)\gg1$), the integrals can be usefully estimated.  The
idea is to Taylor-expand $t(\tang)$ near a stationary point (say
$\tang^*$) where of course an ordinary image would be located.  Near a
stationary point there is no linear dependence on $\tang-\tang^*$ and
the quadratic terms are given by the Hessian (\ref{eq:maginv}) matrix.
We can assume, without loss of generality, that the Hessian is
diagonal.  Let us introduce
\begin{equation}
  a,b = 1 - \kappa \pm\gamma \qquad
  \mu = 1/(ab)
\end{equation}
and then $\mu$ will be the scalar magnification.  In terms of these we
have
\begin{equation}
  t_L(\tang) = t(\tang^*) + {\textstyle\frac12}\Tds
  \left( a(\theta_x-\theta^*_x)^2 + b(\theta_y-\theta^*_y)^2 \right)
\end{equation}
Now both the integrals in (\ref{eq:mag-quotient}) have the form
\begin{equation}
  \int_{-\infty}^\infty \exp\left(\frac{\pm iu^2}{2\sigma^2} \right) \, du
  = \frac{e^{\pm i\pi/4}}{\sqrt{2\pi}\,\sigma}
\end{equation}
Note changing the sign in integrand changes the phase of the integral
by $\pi/2$.  We need to pay attention to the signs of $\omega$, $a$,
and $b$.  Bearing this mind and simplifying, we get
\begin{equation}
  |\mu|^{1/2} \exp\left(i\omega\,t(\tang^*) - i\Delta\right)
\end{equation}
where
\begin{equation}
  \Delta = {\textstyle\frac12}\pi \sgn \omega \times
  \begin{cases}
    \hbox{0 for minima} \\
    \hbox{1 for saddle points} \\
    \hbox{2 for maxima}
  \end{cases}
\end{equation}
The $\Delta$ term here is known as the Morse phase.

\end{document}